\documentclass[prx,twocolumn,groupedaddress]{revtex4}
\usepackage{graphicx,color}
\usepackage{amssymb}   
\usepackage{amsmath}
\usepackage{epstopdf}
\usepackage{natbib}
\usepackage{hyperref}
\usepackage{bm}
\usepackage{epstopdf}
\usepackage{cases}

\newcommand{\bc}{{\bf c}}

\newcommand{\bx}{{\bm x}}
\newcommand{\by}{{\bm y}}
\newcommand{\bp}{{\bm p}}
\newcommand{\bq}{{\bm q}}
\usepackage{color}
\begin{document}


\title{Model for Metal-Insulator Transition in Graphene Superlattices and Beyond}

\author{Noah F. Q. Yuan and Liang Fu}
\affiliation{Department of Physics, Massachusetts Institute of Technology, Cambridge,
Massachusetts 02139, USA}

\begin{abstract}
We propose a two-orbital Hubbard model on an emergent honeycomb lattice to describe the low-energy physics of twisted bilayer graphene. Our model provides a theoretical basis for studying metal-insulator transition, Landau level degeneracy lifting and unconventional superconductivity that are recently observed.
\end{abstract}

\maketitle

The recent discovery of correlated insulator state \cite{Cao1} and unconventional superconductivity \cite{Cao2} in bilayer graphene with a small twist angle has generated tremendous excitement \cite{Mele,Xu,Volovik}. At small twist angles, the moir\'e pattern creates a superlattice with a large unit cell comprising more than $10000$ atoms, and dramatically modify the low-energy electronic structure. In particular, near certain ``magic'' twist angles,  four lowest-energy minibands with a total band width on the order of 10meV are separated from excited bands and accommodate a range of carrier densities from charge $-4e$ to $4e$ per supercell. Due to the strong suppression of kinetic energy in these narrow bands, Coulomb interaction may drive correlated electron phenomena \cite{MacDonald}. Remarkably, the recent experiments \cite{Cao1,Cao2} on such twisted bilayer graphene (TBG) discovered metal-insulator transition and superconductivity at low temperature by tuning the carrier density, applying the magnetic field or slightly varying the twist angle. These fascinating phenomena show a number of similarities with that of cuprates. Notably, a correlated insulating state occurs below $4$K  at the filling of charge $\pm 2e$ per supercell. Under electrostatic doping, two superconducting domes appear on both sides of the insulating state, with a maximum transition temperature $T_c=1.7$K and a record-low carrier density of a few $10^{11}$cm$^{-2}$. The mechanism of metal-insulator transition and superconductivity, the nature of the correlated insulating state and the superconducting state are all open questions.

The Hubbard model is the standard model to study metal-insulator transition driven by the competition of kinetic energy and Coulomb interaction \cite{MIT}. It is also believed to capture key features of the cuprate superconductors \cite{Simons,Stanford, Stripe}. To study metal-insulator transition in TBG, it is highly desirable to identify the real-space Wannier orbitals for the low-energy miniband and find the corresponding tight-binding and Hubbard model.

This is however a nontrivial task that has not been accomplished so far. Thanks to   extensive studies using various methods \cite{key,TB1,TB2,TB3,TB4,TB5, DFT1,DFT2,DFT3, Mele2,MacDonald,Neto,Neto2,Koshino0}, the band structure of TBG at small twist angle is known to be rather complex and depend sensitively on microscopic details such as lattice relaxation. Near the so-called magic twist angle, various methods find four nearly-flat minibands at low energy, but differ significantly on important features such as their bandwidth and the gap to excited bands. Therefore, as a first step,
it is important to extract robust and universal features of these narrow minibands from both theoretical calculation and experimental findings on bilayer graphene at small twist angles.

\begin{figure}
\begin{center}
\leavevmode\includegraphics[width=01.0\hsize]{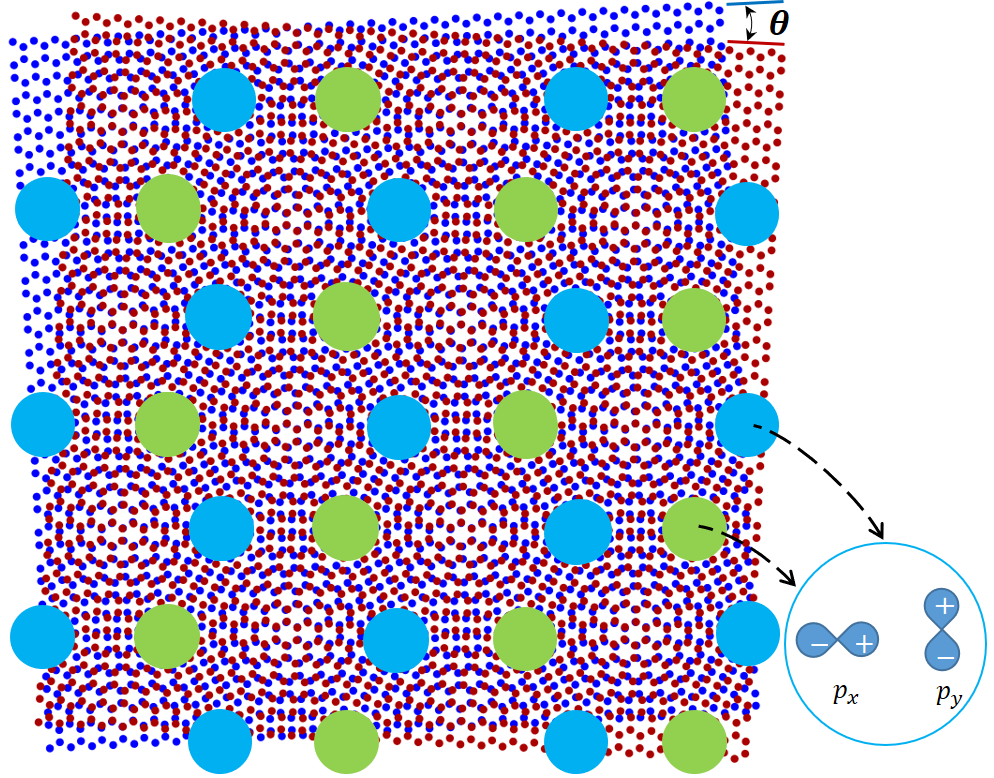}
\end{center}
\caption{Atomic structure and tight-binding model in twisted bilayer graphene. In this figure rotation angle $\theta = 6.01^\circ$. Blue and red little dots represent the carbon atoms of bottom and top layers, respectively. Green and blue giant dots represent AB and BA spots, which form a honeycomb lattice, and AA spots lie in hexagon centers of the honeycomb lattice. At each giant dot (AB/BA spots), there reside two degenerate orbitals with $ p_{x,y} $ symmetries under on-site three-fold rotation.}
\label{fig0}
\end{figure}

In this work, we demonstrate that the electronic structure of narrow minibands and the effect of Coulomb interaction in twisted bilayer graphene are essentially captured by a two-orbital Hubbard model, constructed from Wannier orbitals that extend over the size of supercells. We deduce the centers and symmetry of Wannier orbitals by a straightforward symmetry analysis, without explicitly computing their wavefunctions.  Importantly, the centers of these Wannier orbital form an emergent {\it honeycomb} lattice. The two types of sublattice sites of this honeycomb lattice correspond to AB and BA regions of twisted bilayer graphene respectively, while the hexagon centers correspond to AA regions, as depicted in Fig. \ref{fig0}. At every site of the honeycomb lattice, there are two degenerate Wannier orbitals with $p_x$- and $p_y$-like symmetries, forming a doublet under on-site three-fold rotation.  We then construct an effective tight-binding model on this honeycomb lattice, which reproduces key features of the miniband structure of TBG \cite{key,TB1}. By including Coulomb repulsion, our model provides a useful theoretical basis for studying the metal-insulator transition in TBG as a function of twist angle and carrier density, Dirac fermion reconstruction at charge neutrality, as well as other strongly correlated phenomena such as unconventional superconductivity.

\begin{figure}
\begin{center}
\leavevmode\includegraphics[width=2in]{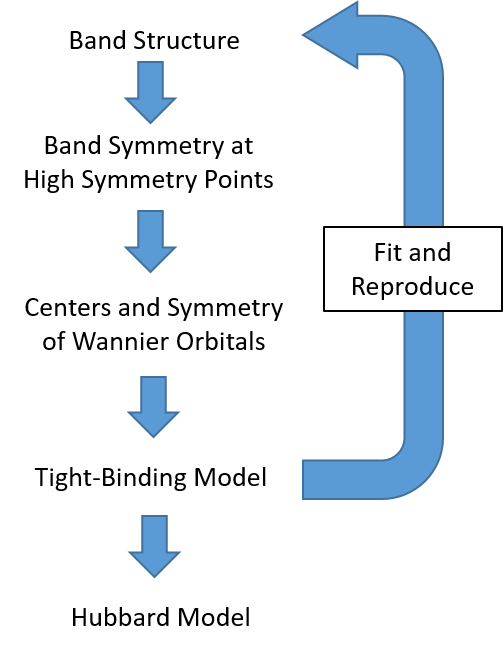}
\end{center}
\caption{The strategy and organization of this work.}
\label{fig00}
\end{figure}

The procedure of our analysis is outlined  in Fig. \ref{fig00}. First, from general considerations on the lowest minibands of TBG and tight-binding calculation by Nam and Koshino \cite{key}, we infer the band symmetry eigenvalues at all high symmetry points $ \Gamma ,K,M $ of mini Brillouin zone (MBZ) of TBG.
Then we examine all possible positions of Wannier centers (which form a lattice in real space) and symmetries of Wannier orbitals ($s$-, $p$-wave etc) to search for solutions consistent with the band symmetries in $k$-space. Luckily, we find that the band symmetries at all high symmetry points can only be reproduced when Wannier orbitals with $ (p_{x},p_{y}) $ on-site symmetry are located on a honeycomb lattice. Based on this important result, we construct the simplest tight-binding model that reproduce the key features of band structure, in particular, the warped Fermi surfaces near the miniband edges at $\Gamma$ point.

This work is organized as follows. In Section I, we study the band structure and energy eigenstates of TBG through the group-theoretical approach. From the obtained energy eigenstates, in Section II we deduce the positions of Wannier orbital centers and the symmetries of Wannier orbitals. Based on the Wannier orbitals, we then construct tight-binding model in Section III. Then, we address the Hubbard model and metal-insulator transition in Section IV, Dirac fermion reconstruction and Landau level degeneracy breaking in Section V, and make connections between theoretical and experimental results. At last in Section VI, we discuss some open questions such as the nature of correlated insulating and unconventional superconducting phases of TBG.

\begin{figure}
\begin{center}
\leavevmode\includegraphics[width=2in]{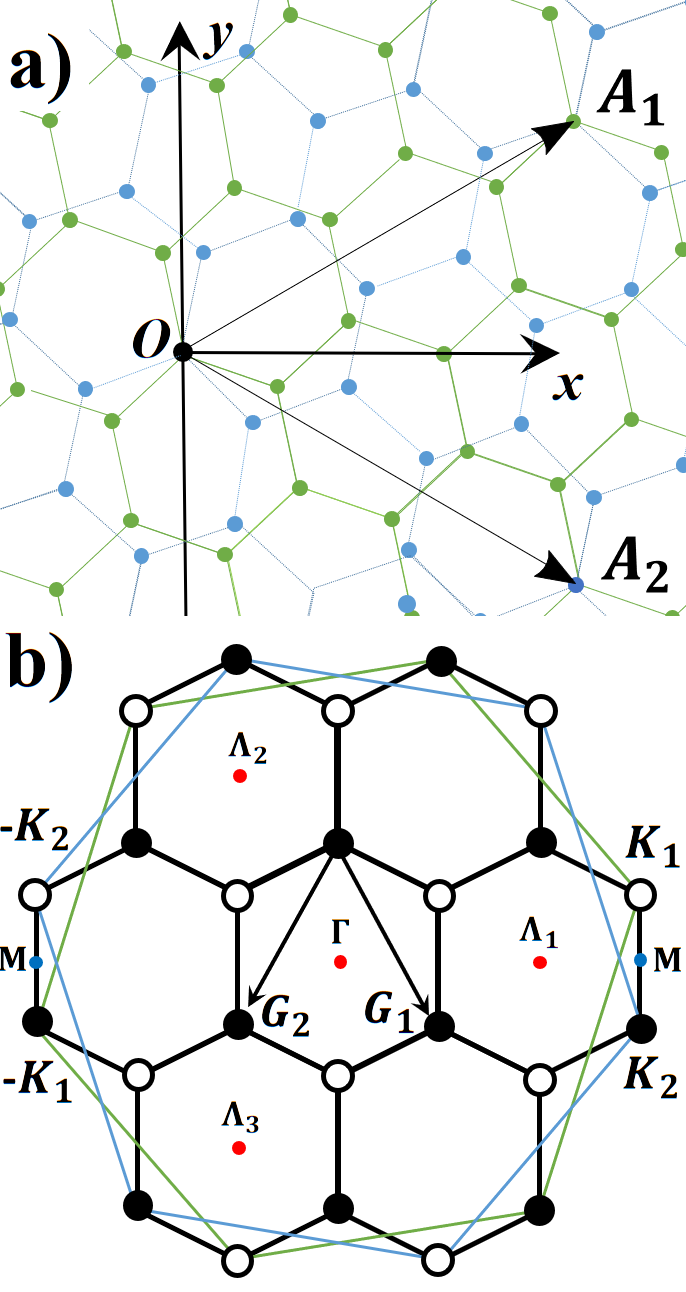}
\end{center}
\caption{ Twisted bilayer graphene with rotation angle $\theta = 21.8^\circ$. a) Atomic structure. Blue and green lines represent the lattices of bottom and top layers, respectively.
b) Mini Brillouin zone (MBZ). Blue and green large hexagons correspond to the first Brillouin zone of bottom and top layers, respectively, and thick small-hexagon to the MBZ. In MBZ, open and filled circles are two inequivalent $K$ points, red dots are equivalent points of $\Gamma$ point, and blue dots are equivalent points of $ M $ point.}
\label{fig_atom1}
\end{figure}

\section{Band Structure}\label{I}
The first question we ask is: are the lowest bands on the electron and hole sides separated from excited bands? Only when low- and high-energy bands are separated by a band gap, a theoretical description using a small number of low-energy degrees of freedom is possible.

To answer this question, we look into theoretical calculation and experimental evidence on the band structure of TBG. Different numerical methods all show that band edges of the lowest-energy electron and hole minibands near charge neutrality are located at $\Gamma$ point of the MBZ. However, conflicting results are found on the gap at $\Gamma$ point between these lowest bands and higher-energy bands when the twist angle is small. Some numerical calculations report relatively large gaps of $10-20$meV on both electron and hole sides \cite{key}, while others show that the gap only exists on the electron side \cite{DFT1} or even no gap exists \cite{TB2,TB4,DFT2}.

In the experiment performed on TBG near the twist angle $\theta =  1.08^{\circ}$, the conductance is found to be zero over a wide range of electron or hole densities near $n=\pm n_s = \pm  2.7 \times 10^{12}$cm$^{-2}$---the density at which the four lowest bands are completely filled or empty. The measured thermal activation gaps are about 40meV, comparable to the single-particle band gaps found by Nam and Koshino in tight-binding calculation with relaxed lattice structure \cite{key}. Hence we conclude that at small twist angles, the lowest bands of TBG are well separated from excited bands above and below.

Our next goal is to understand the band structure and Bloch wavefunction of the lowest minibands analytically. Previous analytical calculations have mostly focused on the minibands near the corners of MBZ \cite{MacDonald,Neto}, where the original Dirac points of graphene remain present, but exhibit a much reduced Fermi velocity. However, a full analysis of energy bands at all high symmetry points remains lacking. 

To do this, we need to work out energy eigenstates of TBG, which are superpositions of states on two layers hybridized through interlayer tunneling.  In real space, the interlayer tunneling takes the general form
\begin{eqnarray}\label{h1}
\mathcal{H}_{\text{T}}=\sum_{m, n=A,B} \sum_{{\bm x}^m,{\bm y}^n}  \xi^{\dagger}({{\bm x}^m}) T({\bm x}^m,{\bm y}^n) \eta({\bm y}^n)+h.c.
\end{eqnarray}
where $\bx^m, \by^n$ denote the coordinates of carbon atom sites on layer 1 and 2 respectively and $m,n$ denotes the $A/B$ sublattice. $\xi$ and $\eta$ are the electron annihilation operators on the two layers respectively.
$T$ describes the tunneling amplitudes between two sites $\bx$ and $\by$ on different layers.

We define electron operator at a given momentum $ \bm k $ as follows:
$ \xi_{\bm k}=(\xi_{\bm k}^{A},\xi_{\bm k}^{B})^{\text{T}},\eta_{\bm k}=(\eta_{\bm k}^{A},\eta_{\bm k}^{B})^{\text{T}} $ where
\begin{eqnarray}
\xi^n_{\bm k} =   \sum_{{\bm x}^n} e^{i {\bm k} \cdot {\bm x}^n} \xi(\bx^n),   \; \;
\eta^n_{\bm k} =   \sum_{{\bm y}^n} e^{i {\bm k} \cdot {\bm y}^n} \eta(\by^n)
\end{eqnarray}
with $n=A, B$. By employing the two-center approximation $ T({\bm x},{\bm y})=T({\bm x}-{\bm y}) $ \cite{MacDonald,Neto,Neto2}, we can write Eq (\ref{h1}) in momentum space as
\begin{eqnarray}\label{h22}
\mathcal{H}_{\text{T}}=\sum_{\bm q,\mathcal{G}_1,\mathcal{G}_2}\xi_{\bm q+\mathcal{G}_{1}}^{\dagger}T_{\bm q}(\mathcal{G}_{1},\mathcal{G}_{2})\eta_{\bm q+\mathcal{G}_{2}}+h.c.
\end{eqnarray}
where $ \mathcal{G}_{1,2} $ is the reciprocal vector of layer 1 or 2 respectively, and $ T_{\bm q}(\mathcal{G}_1 ,\mathcal{G}_2) $ is the 2 by 2 interlayer scattering matrix with momentum transfer $ \mathcal{G}_1 -\mathcal{G}_2 $.

To calculate the energy eigenstates explicitly requires the knowledge of the tunneling operator $ T $. However, when the twist angle is commensurate (as we assume throughout this work), $T$ satisfies certain symmetry conditions, hence at high symmetry points of MBZ the low-energy eigenstates can be classified by the irreducible representations of the corresponding symmetry group. Importantly, essential information on these eigenstates and hence Wannier orbitals can be deduced directly from the symmetry representations. In the following three subsections, we will work out the symmetries of the four lowest minibands at all three high symmetry points: $ \Gamma,K,M $ of MBZ respectively.

Before proceeding, we first introduce the coordinate system. As shown in Fig. \ref{fig_atom1}a, the graphene superlattice has a point group $ D_{3} $ with in-plane three-fold rotation and out-of-plane two-fold rotation. We then choose the axis of two-fold rotation as $y$-axis, and the axis of three-fold rotation as $ z $-axis. Thus two-fold and three-fold rotations can be denoted as $ C_{2y} $ and $ C_{3z} $. The center of $ C_{3z} $, denoted as $ O $ in Fig. \ref{fig_atom1}a, is located at a vertically aligned $A$ site, which we define as the origin of coordinate $\bx=\by=\bm 0$. In this coordinate system, the primitive vectors of the graphene superlattice are $ \bm A_{1,2}=A(\frac{\sqrt{3}}{2},\pm\frac{1}{2}) $ where $ A={a}/[{2\sin(\theta/2)}] $ is the superlattice constant, $ a $ is the common lattice constant of both layers and $ \theta $ is the twist angle. Before band folding, the $+K $ point of the top (bottom) graphene layer in its original Brillouin zone is $\bm K_{1,2}=\frac{4\pi}{3a}\left[\cos\left(\frac{\theta}{2}\right),\pm\sin\left(\frac{\theta}{2}\right)\right]$. After band folding, $+\bm K_{1}$ and $- \bm K_2$ fold onto one corner of MBZ of TBG, while $-\bm K_1$ and $+ \bm K_2$ fold onto the other inequivalent corner, as shown in Fig.\ref{fig_atom1}b.

In our coordinate system, the scattering matrix in (\ref{h22}) can now be explicitly written as 
\begin{eqnarray}\label{T}
T_{\bm q}^{mn}(\mathcal{G}_1 ,\mathcal{G}_2)=t^{mn}(\bm q)\exp\left[-i(\mathcal{G}_1\cdot\bm\tau_{1}^{m}-\mathcal{G}_2\cdot\bm\tau_{2}^{n})\right]
\end{eqnarray}
where $ t^{mn}(\bm q) $ is the Fourier transform of $ T({\bm x}^m -{\bm y}^n) $, and
the vectors $\bm\tau^{A,B}_j$ specify the position of A/B sublattice sites within the unit cell of layer $j$, given by
\begin{eqnarray}
\bm\tau^{A}_{1,2}&=&\bm 0,  \nonumber \\
\bm\tau^{B}_{1,2}&=&\frac{a}{\sqrt{3}}\left[\mp\sin\left(\frac{\theta}{2}\right),\cos\left(\frac{\theta}{2}\right)\right].
\end{eqnarray}
We emphasize that  the phase factor in the interlayer tunneling matrix (\ref{T}) comes from the intra-unit-cell part of the Bloch wavefunction of each graphene layer, and must be included to obtain the correct band structure of TBG.

Since the interlayer tunneling term $\mathcal{H}_{\rm T}$ is much smaller than the full bandwidth of graphene layers, the main effect of $\mathcal{H}_{\rm T}$ is to modify the low-energy part of the band structure near the original Dirac points $ \pm\bm K_{1,2} $.
Therefore, to calculate the low-energy band structure of TBG, it suffices to restrict $ \bm q+\mathcal{G}_{1,2} $ in the interlayer tunneling term (\ref{h22}) to be close to $ \pm\bm K_{1,2} $ points. Then, the full Hilbert space comprising a large number of momentum points is truncated to a reduced Hilbert space comprising only a few momentum points. This  approximation is known as the continuum theory of TBG \cite{Neto,Neto2,MacDonald,Koshino0}.

\subsection{$\Gamma$ Point}
We first study the lowest minibands of TBG at $\Gamma$ point of the MBZ. There are in total three pairs of opposite momenta, denoted as $\pm\bm\Lambda_{a=1,2,3}$,  which are located closest to the original Dirac points $\bm K_{1,2}$ of individual graphene layers {\it and} fold onto $\Gamma$ of MBZ. These three $\Lambda$ points are all integer multiples of the two superlattice reciprocal vectors $\bm G_{1,2}=G_{\theta}\left(\pm\frac{1}{2},-\frac{\sqrt{3}}{2}\right),G_{\theta}=\frac{8\pi}{\sqrt{3}a}\sin\frac{\theta}{2} $, and related by three-fold rotation symmetry, as shown in Fig. \ref{fig_atom1}b.

Interlayer tunneling leads to scatterings among $\pm \Lambda$ points on layer 1 and those on layer 2. Among all possible scattering processes, intervalley scattering between $+\Lambda$ points to $-\Lambda$ points requires large momentum transfer, hence is negligibly small when the twist angle is small \cite{MacDonald}. Thus we only consider intravalley scattering within the three $+\Lambda$ points that are all close to the original Dirac points $+\bm K_{1,2}$, and drop the valley index. Since we focus on $+\bm K_{1,2}$ valleys, we only need to consider two out of the four lowest energy eigenstates at $\Gamma$ point, which are denoted by $\Psi_{e}^{\Gamma}$ on electron side and $\Psi_{e}^{\Gamma}$ on hole side. The other two lowest energy eigenstates coming from $ -\bm K_{1,2} $ valleys will be time-reversal partners of $\Psi_{e,h}^{\Gamma}$.

According to our discussion above, to the lowest order approximation, energy eigenstates at $\Gamma$ are in general superpositons of $ \xi^m_{\bm k},\eta^n_{\bm k} $ where $ \bm k=\bm\Lambda_{a}\quad (a=1,2,3) $. While the energy spectrum depends on the form of tunneling operator $T(\bx^m, \by^n)$, we now deduce purely from general symmetry considerations the essential form of these wavefunctions.

Any energy eigenstate at $\Gamma$ must belong to one of the three irreducible representations of the point group $D_3$: the one dimensional $A_1$ representation, the one-dimensional $A_2$ representation, and the two-dimensional $E$ representation.
The two-fold rotation $C_{2y}$ around $y$-axis maps $+\bm K_{1,2}$ to $-\bm K_{2,1}$ respectively, i.e., simultaneously interchanges the two valleys and the two layers. Therefore, its action cannot be represented within the subspace of states around one valley. Thus it suffices to consider the subgroup $C_3$ that acts within $+\Lambda$ states belonging to the $+{\bm K}$ valley.

In the subgroup $ C_{3} $, every representation is labeled by the eigenvalue of angular momentum $ L_{z}$ taking three possible values: $0, \pm 1$. According to group theory, an eigenstate $ \Psi $ with angular momentum $L_{z}$ formed by Bloch states from three $\Lambda$ points can be generally written as
\begin{eqnarray}\nonumber
\Psi^{\Gamma} &=&\sum_{m}\sum_{\bm x^{m}}\alpha_{m}\sum_{a=1}^{3}e^{i\bm\Lambda_{a}\cdot\bm x^{m}+\frac{2i}{3}aL_{z}\pi}\xi(\bm x^{m})\\
&+&\sum_{n}\sum_{\by^{n}}\beta_{n}\sum_{a=1}^{3}e^{i\bm\Lambda_{a}\cdot\bm y^{n}+\frac{2i}{3}aL_{z}\pi}\eta(\bm y^{n})\label{GT}
\end{eqnarray}
where $ \alpha_{m},\beta_{n} $ are complex coefficients.
Without loss of generality in the following we shall consider the energy eigenstate of the conduction mini-band, denoted by $\Psi_{e}^{\Gamma}$. The analysis of the valence mini-band $ \Psi_{h}^{\Gamma} $ is similar.

We first focus on the case of $L_z = +1$. According to the general expression above, $ \Psi^{\Gamma} $ can be rewritten in the following suggestive way
\begin{eqnarray}\label{e}
\Psi^{\Gamma} &=& e^{-i 2\pi/3} \left[ \alpha_{A}  U^{A}(\bm R_{c})+\alpha_{B}  U^{B} (-\bm R_c) \right] \\\nonumber
&+& e^{i 2\pi/3} \left[ \beta_{A} L^{A} (-\bm R_{c}) +\beta_{B}  L^{B} (\bm R_c) \right].
\end{eqnarray}
Here $ \bm R_{c}=A\hat{\bm x}/\sqrt{3} $ is the coordinate of an BA spot closest to the AA spot at the origin, as shown in Fig. \ref{fig0}. $ U $ and $L$ are electron wavefunctions on the upper and lower layers respectively, defined by
\begin{eqnarray}\label{U}
U^{m}(\bm R)&=&\sum_{\bm x^{m}}e^{i\bm K_{1}\cdot\bm x^{m}} f\left(\bm x^{m}-\bm R\right)\xi(\bm x^m),  \\\label{L}
L^{n}(\bm R)&=&\sum_{\bm y^{n}}e^{i\bm K_{2}\cdot\bm y^{n}}f^{*}\left(\bm y^{n}-\bm R\right)\eta(\bm y^n).
\end{eqnarray}
Both $ U,L $ are the product of intra-unit-cell wavefunction $e^{i\bm K\cdot\bm r}$ which is fast oscillating and the envelope function $f(\bm r-\bm R)$ given by
\begin{eqnarray}\label{env}
f(\bm r)=e^{i(\bm K_{1}-\bm K_{2})\cdot\bm r}\times\left\{1+e^{i\bm G_{1}\cdot\bm r}+e^{i\bm G_{2}\cdot\bm r}\right\},
\end{eqnarray}
which is slowly varying. $ f(\bm r) $ is invariant under three-fold rotation around the origin.

It is straightforward to show that the maxima of the envelop function $|f(\bm r)|$ are located at AA spots $ n_{1}\bm A_{1}+n_{2}\bm A_{2}\quad (n_{1,2}\in\mathbb{Z}) $, which form a triangular lattice with primitive vectors $ \bm A_{1,2} $. Then, the maxima of $|f(\bm r \pm \bm R_c)|$ are located at the AB/BA spots $ \mp \bm R_{c}+n_{1}\bm A_{1}+n_{2}\bm A_{2}$, which also form a triangular lattice but shifted off the origin by $\mp \bm R_{c}$. Therefore, it follows from Eq. (\ref{e}) that  the component of Bloch wavefunction $\Psi^\Gamma$ on the $A$ sublattice has its maxima at BA spots on layer 1 and at AB spots on layer 2, while the component on the $B$ sublattice has its maxima on AB spots on layer 1 and BA spots on layer 2.

This feature is robust and can be understood by symmetry considerations. For a state carrying finite angular momentum such as $L_z=+1$ considered here, its wavefunction necessarily vanishes at rotation centers $n_{1}\bm A_{1}+n_{2}\bm A_{2}$, where $A$ sublattice sites on two layers are registered. Therefore, the maxima of $A$-sublattice component of $\Psi^\Gamma$ must be away from these AA spots. Furthermore, if there is only a single maximum within a supercell (as is the case here), this maximum can only be located at either AB or BA spots, because these positions are invariant under three-fold rotation with respect to AA spots up to superlattice translations. Finally, we note that under the combination of two-fold rotation $ C_{2y} $ and time-reversal symmetry $\mathbb{T}$, the two layers are interchanged, while the sublattice and angular momentum $L_z$ quantum numbers are unchanged.  This implies that the maxima of $A$ sublattice wavefunction is symmetric under $ C_{2y}$, hence must be located at BA spots on layer 1 and AB spots on layer 2, forming the $C_{2y}$ image of each other.
On the other hand, $B$ sublattice sites do not coincide with rotation centers, and the intra-unit-cell phase factors $e^{i {\bm K}_1 \cdot {\bm x}^B}$ and $e^{i {\bm K}_2 \cdot {\bm y}^B}$ already carry the angular momentum $ -1 $. To make the total angular momentum $L_z=+1$, the envelope function of $B$ sublattice wavefunction must carry $ -1 $ angular momentum and is therefore peaked at AB spots on layer 1 and BA spots on layer 2. Here we have used the fact that under $ C_{3z} $, angular momentum is defined modulo 3.

We thus conclude that, to respect the point group $ D_{3} $ of twisted bilayer graphene, the energy eigenstate at $\Gamma$ point with angular momentum $ L_{z}=+1 $ has the peculiar property that $A$ or $B$ sublattice wavefunction has its maxima on a honeycomb lattice whose two sublattices correspond to AB and BA spots of the two layers. Similar conclusion is found for eigenstates with $ L_{z}=-1 $, which we leave to the Supplementary Material \cite{SM}.

In the band structure calculations by Nam and Koshino \cite{key}, at $\Gamma $ point the eigenstates form two doublets---one on electron side and one on hole side, and each doublet splits along $\Gamma M$ line in MBZ. It is our understanding that such band structure is consistent with the scenario that the two members of the doublet come from $ \pm\bm K$ valleys of graphene and have $ L_{z}=\pm 1 $ respectively. In this case, the two-fold degeneracy is protected by $D_3$ point group and time-reversal symmetry and remains intact even when inter-valley scattering is taken into account.
In contrast, in the scenario of $ L_{z}=0 $, there will be four non-degenerate states at $\Gamma$, which is not found in band structure calculations. For completeness, we leave the discussion of $ L_{z}=0 $ case to the Supplementary Material \cite{SM}.

So far our analysis has been completely based on symmetry considerations. To work out the eigenstates $ \Psi_{e,h}^{\Gamma} $ at $\Gamma$ point explicitly requires a microscopic calculation, which we sketch below. Before turning on interlayer tunneling, the Hamiltonian for graphene states at $\Lambda$ points is given by
\begin{eqnarray}\label{h2}
H_0 =
 v_{F}\sum_{a=1}^3 [\xi^{\dagger}_{a}{(\bp_{a}} \cdot \bm \sigma^{1})\xi_{a} +  \eta^{\dagger}_{a} {(\bq_{a}} \cdot \bm \sigma^{2})\eta_{a}].
\end{eqnarray}
Here $ \xi_a$ and $\eta_a$ are respectively the electron operator of layer 1 and 2 at momentum $\bm\Lambda_a$. In $H_0$, $\bp_a$ and $\bq_a$ are the small momentum differences between $\Lambda_a$ and the Dirac points on layer 1 and 2 respectively, and $ v_F $ is the Fermi velocity.
Due to their difference in orientation, the Pauli operators in the Dirac Hamiltonian of individual graphene layers are ``rotated'' oppositely:
\begin{eqnarray}
{\bm \sigma}^{1,2} ={\Big (}   \sigma_{x}\cos\frac{\theta}{2} \mp \sigma_{y}\sin\frac{\theta}{2}, 
 \pm \sigma_{x}\sin\frac{\theta}{2}+\sigma_{y}\cos\frac{\theta}{2} {\Big )},
\end{eqnarray}
where $ \sigma_{x,y} $ are Pauli matrices.

According to Eqs.(\ref{h1}) and (\ref{h22}), the interlayer tunneling between $\Lambda$ points can be written as
\begin{eqnarray}\label{h3}
H_{\text{T}}=\sum_{a,b=1}^{3}\xi_{a}^{\dagger}T_{ab}\eta_{b}+h.c.
\end{eqnarray}
where $ T_{ab} $ is the scattering matrix obtained by integrating out intermediate high-energy states in Eq.(\ref{h22}) other than the low-energy $\pm \bm \Lambda_{a}$  states kept in Eq.(\ref{h3}).

The three-fold rotation of $\Lambda$ points is realized by the cyclic permutation $ \Lambda_{a}\to\Lambda_{a+1} $ (with $ \Lambda_4 \equiv \Lambda_1 $).
Due to the invariance of $ H_{\text{T}} $ under three-fold rotation and with a proper choice of relative phases of $\xi_a$ and $\eta_a$, we have $T_{a+1,b+1}=T_{ab}$. Hence essentially there are only three independent scattering matrices among all of them, and we denote them as $ T_{0}=T_{aa},T_{\pm}=T_{a\pm 1,a} $. Physically, interlayer tunneling associated with $ T_{0} $ preserves momentum while $ T_{\pm} $ transfers momentum $ \pm(\bm G_{1}-\bm G_{2}) $ from layer 2 to layer 1. 

\begin{figure}
\begin{center}
\leavevmode\includegraphics[width=01.0\hsize]{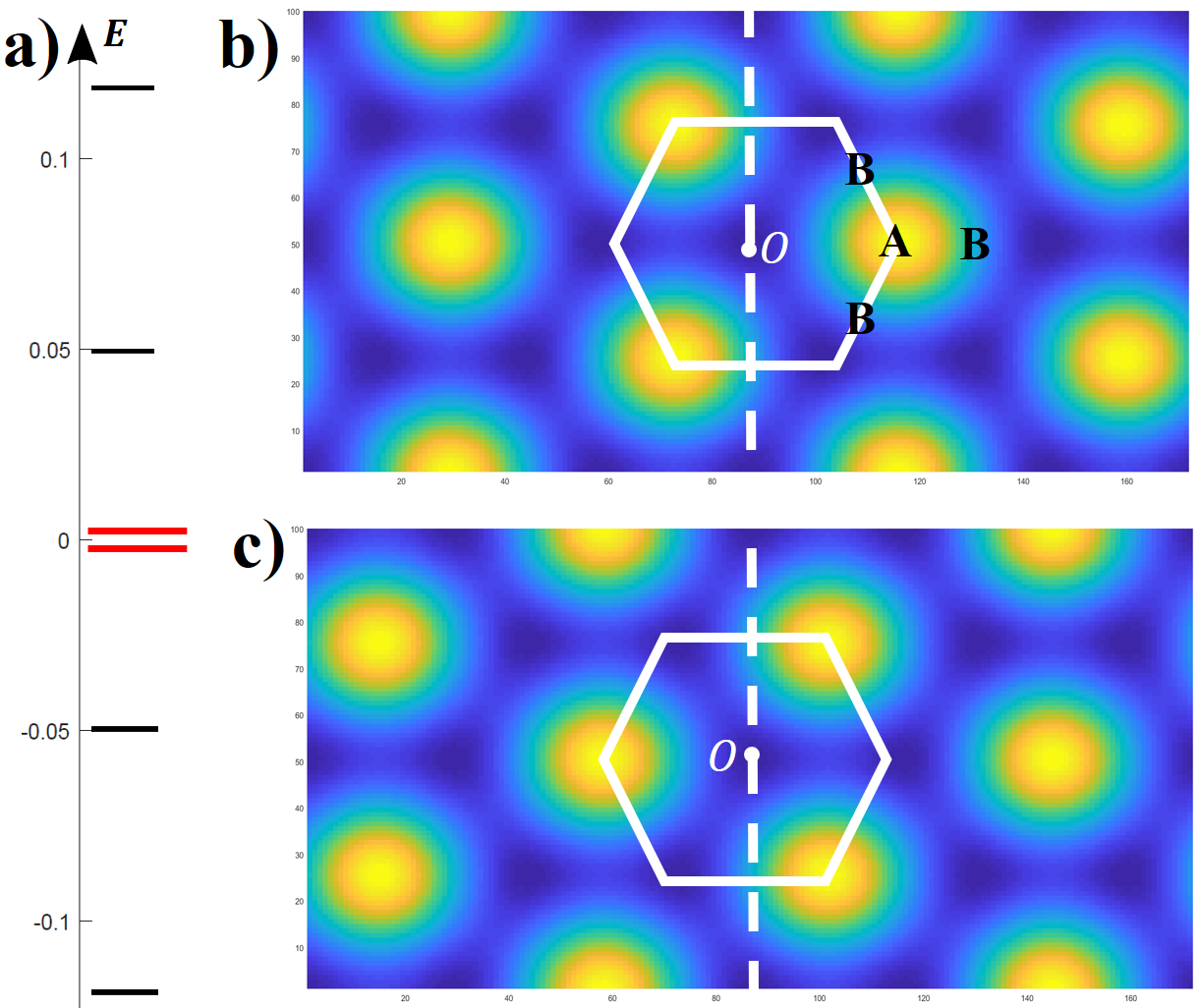}
\end{center}
    \caption{Real space probability distributions $ |\Psi_{e}^{\Gamma}(\bm x)|^2=|\Psi_{h}^{\Gamma}(\bm x)|^2 $ of lowest eigenstates at $\Gamma$ point with angular momentum $ L_{z}=+1 $. Eigenenergies are shown in a) in unit of $2v_{F}/(\sqrt{3}a)$, where the lowest energies are in red. $ |\Psi_{e,h}^{\Gamma}(\bm x)|^2 $ in layer 1 and 2 are shown in b) and c) respectively. The hexagon is the unit cell of graphene superlattice and $ O $ is origin. Details can be found in the Supplementary Material \cite{SM}.
}
\label{3}
\end{figure}

With (\ref{h2}) and (\ref{h3}), we can work out the lowest eigenstates $ \Psi_{e}^{\Gamma} $ and $ \Psi_{h}^{\Gamma} $ at $ \Gamma $ point. An illustrative example of $ \Psi_{e}^{\Gamma} $ and $ \Psi_{h}^{\Gamma} $ with $ L_z =+1 $ is shown in Fig. \ref{3} where $A$ sublattice component dominates and its maxima are indeed located at the emergent honeycomb lattice formed by AB and BA spots. Details of interlayer coupling Hamiltonian (\ref{h3}) in this example can be found in the Supplementary Material \cite{SM}.

\subsection{K points}
Similar analysis as shown above applies to energy eigenstates at other high symmetry points in the MBZ. Consider the corners of MBZ, denoted by $\pm \bm K$ where $ \bm K\equiv\bm K_{1}-\bm K_{2} $. Since time-reversal symmetry relates bands at $\pm \bm K$ by complex conjugation, it suffices to study $+\bm K$ only. Since $+\bm K_1$ and $-\bm K_2$ in the large Brillouin zone fold onto $+\bm K$ in the MBZ, low-energy states of TBG at $+\bm K$ come predominantly from the states of layer 1 at Dirac point $+\bm K_1$ (denoted by $\xi^{A,B}_+$) and the states of layer 2 at Dirac point $-\bm K_2$ (denoted by $\eta^{A,B}_-$). When the intervalley scattering is negligibly small, to leading order approximation we obtain four degenerate zero-energy states at $+\bm K$, as found in previous studies.

At $\bm K$ point, the point group $D_{3}$ is preserved and hence the eigenstates can also be labeled by representations $A_{1},A_{2},E$ of $D_3$. Under three-fold rotation, we find that states at $ A $ sites of each layer have zero angular momentum, and $\xi^{A}_+ +\eta^{A}_-$ will furnish the trivial representation $A_1$ while $\xi^{A}_+ -\eta^{A}_-$ will furnish the 1D representation $A_2$. The remaining two states $(\xi^{B}_+ ,\eta^{B}_-)$ at $B$ sites have finite, opposite angular momenta. Furthermore, under two-fold rotation $ C_{2y} $, the type of sublattice is unchanged but $ +\bm K_{1} $ and $-\bm K_{2}$ are interchanged. As a result, $(\xi^{B}_+ ,\eta^{B}_-)$ furnish the 2D representation $E$ of $D_3$.

Previous studies found that the correction to the above energy eigenstates from excited states of individual graphene layers will not lift the degeneracy at Dirac point but only reduces the Fermi velocity. We shall come back to this point later.

\subsection{M points}
At $M$ points, only in-plane two-fold rotation $C_{2y}$ is present. Thus one can classify eigenstates at $M$ points according to the eigenvalues $\pm 1$ of $C_{2y}$. Energy eigenstates at $M$ points are from the state with momentum $ \pm\frac{1}{2}(\bm K_1 +\bm K_2) $ on both layers. Since $ C_{2y} $ interchanges the two layers and the two momenta, we obtain two energy eigenstates with $ C_{2y}=+1 $ and two with $ C_{2y}=-1 $.

The representations of the eigenstates at high symmetry points $ \Gamma,K,M $ are summarized in Table. \ref{S}.

\section{Wannier Orbitals and Centers}
In this section we deduce the center and symmetry of Wannier orbitals from symmetries of band eigenstates at high symmetry points through consistency check. While the detailed form of Wannier orbitals are to be obtained by Fourier transform of band eigenstates over the MBZ, the center and symmetry of Wannier orbitals are robust features that only depend on symmetries of energy bands at all high symmetry points.

Among all possible configurations of Wannier orbitals, we first consider the case of degenerate $ (p_x ,p_y) $-like Wannier orbitals at the honeycomb lattice formed by AB and BA spots as shown in Fig. \ref{fig0}. This choice is motivated by the property of energy eigenstates $ \Psi^{\Gamma} $ shown earlier. We show below that  the case of $ (p_x ,p_y) $ Wannier orbitals on the honeycomb lattice yields the correct band symmetries at all high symmetry points of MBZ shown in Table. \ref{S}.

To see this explicitly, it is useful to construct the electron Bloch state with momentum $ \bm k\in $MBZ in $ p_{x} \pm i p_{y} $ orbital basis  as follows
\begin{eqnarray}\label{Bloch}
c_{\bm k,\tau}=\sum_{j}e^{i\bm k\cdot\bm R_{j}}c_{j\tau}
\end{eqnarray}
where $ \tau=\pm $ denotes the Wannier orbital with $L_z= \pm 1$ under three-fold rotation around its own center, $ \bm R_{j} $ is the coordinate of honeycomb lattice site $j$, $ c_{j\tau} $ annihilates an electron at site $j$ with orbital $\tau$, and the sum is over all sites of the honeycomb lattice.


At $\Gamma$ point $ \bm k=\bm 0 $, the plane-wave phase factor $ e^{i\bm k\cdot\bm R_j}=1 $ does not contribute angular momentum, and the angular momentum of Bloch  wavefunctions $c_{\bm 0,\tau}$ comes solely from the $p_x \pm i p_y$ on-site symmetry of Wannier orbital $ c_{j\tau} $, leading to the desired angular momentum $ L_{z}=\tau =\pm 1 $. The two-fold rotation $ C_{2y} $ interchanges layer indices 1 and 2, and maps $ p_{x}+ip_{y} $ to $ -p_{x}+ip_{y} $ and vice versa. As a result, the $ p_{x}+ip_{y} $ Wannier orbital at AB spot and the $ p_{x}-ip_{y} $ orbital at BA spot will furnish 2D representation $E$ of the $D_3$ point group. The same holds for the $ p_{x}+ip_{y} $ orbital at BA spot and $ p_{x}-ip_{y} $ orbital at AB spot. This leads to two doublets as energy eigenstates at $\Gamma$ point, consistent with our results in Table. \ref{S}.

At $\bm K$ point $ \bm k=\bm K_{1}-\bm K_{2} $, 
the plane-wave phase factor $ e^{i\bm k\cdot\bm R_j}$ on the honeycomb lattice generates nonzero angular momentum, which adds to the orbital angular momentum of $p_x \pm i p_y$ Wannier orbitals. To see this explicitly, we decompose the Bloch state (\ref{Bloch}) as the superposition of Bloch waves in different sublattices
\begin{eqnarray}
c_{\bm k,\tau}=c_{\bm k,\tau,1}+c_{\bm k,\tau,2},\quad c_{\bm k,\tau ,n}\equiv\sum_{j\in L_{n}}e^{i\bm k\cdot\bm R_{j}}c_{j\tau},
\end{eqnarray}
where $ L_{1,2}=\{\mp\bm R_{c}+n_{1}\bm A_{1}+n_{2}\bm A_{2}|n_{1,2}\in\mathbb{Z}\} $ denote the set of AB/BA spots respectively.

We then find under the action of $C_{3z}$ around the origin,
\begin{eqnarray}\nonumber
c_{\bm K,\tau,1}&\to &
\sum_{j\in L_{1}}e^{i\bm K\cdot C_{3z}\bm R_{j}}e^{-2i\tau\pi/3}c_{j\tau}\\\nonumber
&=&\sum_{j\in L_{1}}e^{i\bm K\cdot\bm R_{j}+2i\pi/3}e^{-2i\tau\pi/3}c_{j\tau}\\
&=&e^{-2i(\tau -1)\pi/3}c_{\bm K,\tau,1}
\end{eqnarray}
and hence the Bloch state $ c_{\bm K,\tau,1} $ has angular momentum $ \tau -1 $ with an additional $ -1 $ angular momentum from Bloch wave phase factor $ e^{i\bm k\cdot\bm R_j} $. Likewise, the Bloch state $ c_{\bm K,\tau,2} $ has angular momentum $ \tau +1 $ under three-fold rotation $C_{3z}$. With $ \tau=\pm 1 $ we find the four Bloch states $ c_{\bm K,\tau,n} $ fall into two categories: one pair of doublet $ (c_{\bm K,-,1},c_{\bm K,+,2}) $ and two singlets $ c_{\bm K,+,1},c_{\bm K,-,2} $. Notice that the two-fold rotation $ C_{2y} $ maps $ c_{\bm K,\tau,1} $ to $ -c_{\bm K,-\tau,2} $ and vice versa, we conclude that the doublet $ (c_{\bm K,-,1},c_{\bm K,+,2}) $ furnishes the 2D representation $E$, while the singlets $ c_{\bm K,+,1}\mp c_{\bm K,-,2} $ furnish the 1D representation $ A_{1,2} $ respectively. The case with $ -\bm K $ point can be obtained through time-reversal operation. 

At $ M $ points $ \bm k=\frac{1}{2}(\bm K_{1}+\bm K_{2}) $, the little group is $ C_{2} $. As discussed in section I, the Abelian group $ C_{2} $ can only host 1D representations denoted as $A$ and $B$. Since $ \bm M=\frac{1}{2}(\bm K_{1}+\bm K_{2}) $ is invariant under $ C_{2y} $ up to a reciprocal vector in MBZ, we find $ c_{\bm M,+,1}- c_{\bm M,-,2}$ and $ c_{\bm M,-,1}- c_{\bm M,+,2} $  belong to $ A $ representation while $ c_{\bm M,+,1}+ c_{\bm M,-,2} $ and $ c_{\bm M,-,1}+c_{\bm M,+,2} $  belong to $ B $ representation.

We further show in the Supplementary Material \cite{SM} that from an exhaustive case-by-case study, any other type of lattice or orbital symmetry is incompatible with the band symmetry at all high symmetry points. Thus, we conclude that the Wannier orbitals of TBG have $ p_{x,y} $ symmetries and form a honeycomb lattice whose sites correspond to AB and BA spots.

\begin{table}
\centering
\begin{tabular}{c|ccc} \hline
         & $\Gamma$          & $K$          & $M$     \\
            \hline
Group    & $  D_3 $             & $D_3$            & $C_{2}$ \\
Reps     & $ \{E,E\} $      & $\{A_{1},A_{2},E\}$            & $\{A,A,B,B\}$ \\
$C_{3z}$ & $\{\sigma_{z},\sigma_{z}\}$   & $\{0,0,\sigma_{z}\}$    &  NA     \\
$C_{2y}$ & $\{\sigma_{x},\sigma_{x}\}$   & $\{+,-,\sigma_{x}\}$   &$\{+,+,-,-\}$\\
\hline
\end{tabular}
\caption{Symmetries of lowest four eigenstates at $ \Gamma,K,M $ points. At each high symmetry point, the first line denotes the symmetry group (Group). The lowest four bands furnish irreducible representations (Reps) shown in second line, which can be labeled by representations of group elements $ C_{3z},C_{2y} $ shown in last two lines. The representation of $ C_{3z}=e^{-i\frac{2\pi}{3} L_{z}} $ is denoted by the angular momentum $ L_z $, and Pauli matrix $ \sigma_{z} $ denotes a pair of doublet with $ L_z=\pm $. Since $ C_{3z} $  is not in group $C_{2}$, eigenvalues of $C_{3z}$ are not applicable (NA) in $C_2$. Since $ C_{3z} $ and $ C_{2y} $ do not commute, when $ L_{z}\neq 0 $, the representation of $C_{2y}$ will be Pauli matrix $ \sigma_{x} $. }
\label{S}
\end{table}

The exact form of Wannier orbitals of TBG are to be constructed from energy eigenstates at all momenta in the MBZ, which we leave for a future work. We expect that these Wannier orbitals will extend over multiple supercells of TBG.

\section{Tight-Binding Model}

In this section we construct a tight-binding model on the honeycomb lattice with $ (p_{x},p_{y}) $-like orbitals as an effective model for the four lowest bands of TBG. Our construction is guided by the $D_3$ point group symmetry of TBG and the  valley $U(1)$ symmetry that emerges in the absence of inter-valley scattering.

As is well-known, $p$-orbitals are allowed to have two types of hopping, $\sigma$ (head-to-tail) and $\pi$ hopping (shoulder-to-shoulder). When the $\sigma$ and $\pi$ hopping amplitudes are the same, we arrive at the simplest tight-binding model of $p$-orbitals on the honeycomb lattice
\begin{eqnarray}\label{TB0}
H_{0}=-\sum_{i}\mu \bc^\dagger_{i}\cdot\bc_{i}+\sum_{\langle ij\rangle}    t_1 [\bc^\dagger_{i}  \cdot\bc_{j} + h.c.]\\\nonumber
+\sum_{\langle ij\rangle'} t_{2}[{\bc}^\dagger_{i}\cdot{\bc}_{j}+ h.c.]
\end{eqnarray}
where $\bc_{i}= (c_{i,x}, c_{i,y})^{\text{T}}$ with $ c_{i,x(y)} $ annihilating an electron with $p_{x(y)}$-orbital at site $ i $. $ t_{1} $ is the real hopping amplitude between nearest neighbor sites of different sublattices. $ t_{2} $ denotes real hopping amplitudes between fifth nearest neighbors with bond length $ \sqrt{3}A $, or equivalently, second nearest neighbors within the same sublattice. With only the $ t_1$ term, the band structure is particle-hole symmetric, while the $t_2$ term breaks this symmetry.
$ \mu $ is the on-site chemical potential. Further neighbor hoppings can be included as well.

\begin{figure}
\begin{center}
\leavevmode\includegraphics[width=1\hsize]{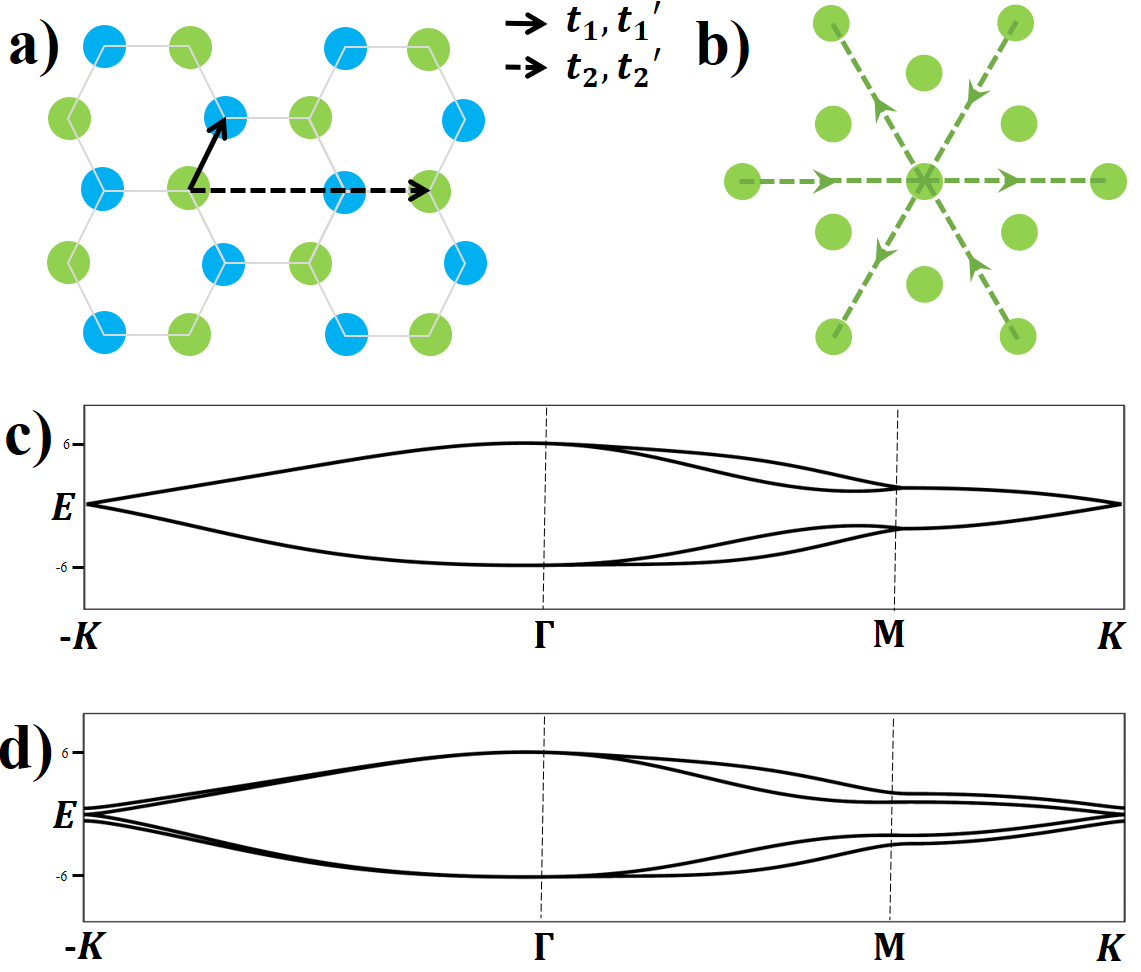}
\end{center}
\caption{a) Hoppings between the nearest neighbors $ t_1 ,t_1' $ and fifth nearest neighbors $ t_2, t_2' $, as shown in Eqs. (\ref{TB0})-(\ref{TB2}). b) Real space representation of (\ref{TB1}) in the green sublattice. Along arrow directions the hopping terms are $ \mp i t'_{2} $ for orbital $ c_{\pm} $. Here we consider hopping terms associated with the central site as an example, and hopping terms associated with other sites can be obtained by lattice translation. The blue sublattice has the same hopping pattern. c) Band structure of the tight-binding model $ H_{tb} $ with $ t_{1}=2,t'_{1}=0,t_{2}=0.05,t'_{2}=0.2,\mu =0 $. d) Band structure with the same parameters in c) except $ t_{1}'=0.2 $.}
\label{fig}
\end{figure}

The Hamiltonian (\ref{TB0}) has emergent SU(4) symmetry and hence every band is four-fold degenerate including orbital and spin. This result however does not match with numerical band structure by Nam and Koshino \cite{key} where the orbital degeneracy is found broken along $ \Gamma M $ lines. To describe such effect, we include the following Hamiltonian
\begin{eqnarray}\label{TB1}
H_{1}&=&\sum_{\langle ij\rangle'}t_{2}'[ ({\bc}^\dagger_{i}\times{\bc}_{j})_{z}+ h.c.]\nonumber\\
&=&-i\sum_{\langle ij\rangle'}t_{2}' ({c}^\dagger_{i+}{c}_{j+}-{c}^\dagger_{i-}{c}_{j-})+ h.c.
\end{eqnarray}
In terms of the chiral basis $ c_{\pm}=(c_{x}\pm ic_{y})/\sqrt{2} $ associated with $ p_{x}\pm ip_y $ orbitals, the hopping terms in (\ref{TB1}) become imaninery values. As illustrated in Fig. \ref{fig}b, orbitals of different chirality experience finite and opposite magnetic fluxes, and the model as a whole preserves time-reversal symmetry. Microscopically the chiral Wannier orbitals $ c_{\pm} $ originate from states near different valleys $ \pm\bm K$ of graphene. When intervalley coupling is negligible, the orbital U(1) symmetry $c_\tau \rightarrow e^{i \tau \phi} c_\tau$ is respected.
$H_1$ breaks SU(4) symmetry down to U(1)$\times $SU(2), where U(1) refers to orbital chirality conservation and SU(2) refers to spin rotation symmetry.

In principle, we can further include the Hamiltonian breaking orbital U(1) symmetry as follows
\begin{eqnarray}\label{TB2}
H_{2} =\sum_{\langle ij\rangle} t'_1 [ \bc^\dagger_{i}  \cdot {\bm e}^{\parallel}_{ij}  \; {\bm e}^{\parallel}_{ij} \cdot  \bc_{j} -\bc^\dagger_{i} \cdot {\bm e}^{\perp}_{ij}  \; {\bm e}^{\perp}_{ij} \cdot  \bc_{j}+ h.c.]
\end{eqnarray}
where $\bm e^{\parallel,\perp}_{ij}$ denote in-plane unit vectors in the direction parallel and perpendicular to the nearest neighbor bond $\langle ij\rangle$ respectively. This symmetry-breaking term arises when intervalley scattering is included, and leads to Dirac mass generation which we will discuss later in Section V.

Symmetry-breaking terms (\ref{TB1}) and (\ref{TB2}) have important effects on the band structure. Denoting $\bm\tau$ as Pauli matrices acting in chiral orbital index and $ \bm\sigma $ as Pauli matrices acting in the sublattice index of the honeycomb lattice, then in the chiral basis the tight-binding Hamiltonian $ H_{tb} $ leads to the $ k\cdot p $ Hamiltonian near $\Gamma$ point in MBZ
\begin{eqnarray}
H(\bm k)&=&H_{1}(\bm k)+H_{2}(\bm k),\\\nonumber
H_{1}(\bm k)&=&t_{1}\left(3-\frac{A^2}{4}|\bm k|^2\right)\sigma_{x}
+v\bm k\cdot\bm\tau\sigma_{y}-\mu,\\\nonumber
H_{2}(\bm k)&=&t_{2}\left(6-\frac{9A^2}{2}|\bm k|^2\right)+\lambda(k_+^3 + k_-^3) \tau_z,
\end{eqnarray}
where $v=\frac{\sqrt{3}}{2}At_{1}',\quad\lambda =-\frac{3\sqrt{3}}{8}A^3 t'_{2},\quad k_{\pm}=k_{x}\pm ik_{y}$.

Interestingly, the $t_{2}'$ term $(k_{+}^3+k_{-}^3)\tau_{z}$  is the orbital analog of hexagonal warping in spin-helical surface states of topological insulators \cite{LF}.
In the full MBZ, the hexagonal warping term vanishes along $ \Gamma K $ and $ MK $ lines and becomes finite along $\Gamma M$ lines. Hence orbital degeneracy is preserved by hexagonal warping term along $ \Gamma K $ and $MK$ lines, consistent with numerical band structures \cite{key,Cao3}. The $t_{1}'$ term further beaks the orbital degeneracy at any point of the MBZ except $ \Gamma, K $, which corresponds to the reduced symmetry group $ D_{3}\times $SU(2) where SU(2) is spin rotation group and $ D_3 $ is point group acting jointly on lattice sites and $ (p_x,p_y) $ orbitals. These effects can be seen from band structures in Fig. \ref{fig}c and d where the band structure in Fig. \ref{fig}c with $ t_{1}'=0 $ agrees with calculations by Nam and Koshino \cite{key}, and the band structure in Fig. \ref{fig}d with finite $t_{1}'$ shows no band degeneracy except $ \Gamma,K $ points where $ D_{3} $ is respected and 2D representation is realized.

\section{Hubbard Model and Metal-Insulator Transition}
From numerical calculations on band structure of TBG, we expect the lowest band width and hence hopping parameters $ t_{1},t_{2},t'_{1},t'_{2} $ are of order of meV. With such small kinetic energy, the effect of interaction have to be considered. In terms of Wannier orbitals, we write down the general form of symmetry-allowed two-orbital on-site Hubbard interaction Hamiltonian
\begin{eqnarray}
&&H_{UJ}=U\sum_{i,\tau}n_{i\tau\uparrow}n_{i\tau\downarrow}+U'\sum_{i}n_{ix}n_{iy}\\\nonumber
&+&J\sum_{i,s,s'}c^{\dagger}_{ixs}c^{\dagger}_{iys'}c_{ixs'}c_{iys}
+J'\sum_{i,\tau\neq\tau'}c^{\dagger}_{i\tau\uparrow}c^{\dagger}_{i\tau\downarrow}c_{i\tau'\downarrow}c_{i\tau'\uparrow}
\end{eqnarray}
where $ U,U',J,J' $ denote the intraorbital Coulomb, interorbital Coulomb, exchange, and pair-hopping interactions, respectively. $ \tau=x,y $ is the orbital index and $ s=\uparrow,\downarrow $ is the spin index. Here $ n_{i\tau s}=c_{i\tau s}^{\dagger}c_{i\tau s} $ is the number of electrons at site $i$ with orbital $\tau$ and spin $s$, and $n_{i\tau}=n_{i\tau\uparrow}+n_{i\tau\downarrow} $. 
Since we choose real $ p_{x,y} $ orbitals, we further find $ J=J' $ denotes the Hund's coupling.

In general, the density-density interactions should be comparable with each other but much larger than Hund's coupling, $ U\sim U'\gg J,J' $. At the simplest level of approximation, we set $ U=U' $ and write down the two-orbital Hubbard model as $ H=H_{0}+H_{1}+H_{2}+H_{U} $ where the on-site interaction is
\begin{equation}
H_{U}=\frac{1}{2} U\sum_{i}(n_{i}-1)^2\quad (n_{i}=n_{ix}+n_{iy}),
\end{equation}
and $ H_{0,1,2} $ are given in (\ref{TB0}), (\ref{TB1}) and (\ref{TB2}).

As mentioned earlier in Section II, the Wannier orbitals are expected to extend over the supercell. Hence electron-electron interaction between nearby sites
\begin{eqnarray}
H_{V}=\sum_{ij}V_{ij}n_{i}n_{j}
\end{eqnarray}
is also present and may be important. Finally we note that the strength of Coulomb repulsion is reduced by screening from excited bands of TBG which span a wide range of energies from 10 meV to 10 eV. Furthermore, in typical graphene devices, the long-range Coulomb interaction is screened by nearby metallic gates at a distance comparable to the size of the supercell. We thus expect a model with a few short-range interactions is sufficient.

We now address the metal-insulator transition in the two-orbital honeycomb Hubbard model and connect our findings to experimental results. With two sites per unit cell and two orbitals at every site, our honeycomb lattice model can accommodate up to 8 electrons per unit cell, which corresponds to the complete filling of the miniband in TBG. The charge neutrality point of TBG corresponds to on average $n=$4 electrons per unit cell or equivalently $ q=2 $ electrons per site of honeycomb lattice in our model. The Mott insulator state found in experiments occurs at $n=$2 electrons/holes per unit cell, or equivalently $ q=1 $ electron/hole per site.

Generally speaking, in Hubbard models with an integer average number $ q $ of electrons per site, a transition from metal to Mott insulator is induced by increasing the ratio of Coulomb repulsion and bandwidth $U/t$. When the twist angle approaches the magic value, the miniband bandwidth decreases very rapidly \cite{key,TB2}. Therefore varying the twist angle slightly induces a bandwidth-controlled metal-insulator transition.
Importantly, in two-orbital Hubbard models, the critical value of $U$ depends on $q$. In the SU(4) symmetric limit, the critical coupling $U_{c}$ for $q=2$ is known to be larger than that for $q=1$, because orbital fluctuations are largest in the former case \cite{George, Kotliar}. This is consistent with the experimental finding of the insulating state at $q=1$, but not at $q=2$ (charge neutrality).

Transport measurement reveals that the insulating state in TBG has a thermal activation gap of 0.3meV, and is destroyed by the application of an in-plane or perpendicular magnetic field of 8T corresponding to a Zeeman energy of 0.5meV \cite{Cao1}. This implies that the charge gap of the correlated insulator is much smaller than the bandwidth. Therefore, the insulating state is a weak Mott insulator close to a metal-insulator transition, ruling out the possibility of $U\gg t$. Moreover, before it becomes highly insulating at low temperature,  above 4K resistivity increases linearly with temperature. This behavior is characteristic of high-temperature bad-metal regime of a Mott insulator with or without doping, seen in many oxides and in numerical study of Hubbard model \cite{badmetal}.

\section{Dirac Fermion reconstruction at charge neutrality}

We now highlight a interesting finding in the experimental paper \cite{Cao1}, whose significance may have not been fully recognized.
Over a wide range of densities around charge neutrality, Landau level degeneracy is  found to be $4$-fold instead of $8$-fold, as expected from the 8 degenerate massless Dirac fermions coming from graphene valley, electron spin, and $\pm \bm K$ points of MBZ.

Our work offers a possible solution to this problem.
As we showed previously in Section I and III,
after band folding the Dirac doublet at $+\bm K_{1}$ in layer 1 and the one at $-\bm K_{2}$ in layer 2 will fall on top of each other at $+\bm K$ of MBZ. From symmetry considerations, the resulting four states at $+\bm K$ consist of one doublet ($E$)  and two singlets ($A_1$ and $A_2$) representation of the $D_3$ point group. The same result applies to the $-\bm K$ point. Therefore half of the Dirac fermions in the doublet representation are protected by symmetry, while the other half in the singlet representation are unstable and can become massive if perturbations are considered.

First, the interlayer tunneling Hamiltonian (\ref{h1}) contains terms that scatter states at different valleys on different layers, which generates single-particle mass for the half unstable Dirac fermions. However, as we mentioned earlier,  the intervalley scattering requires large momentum transfer and is very weak when the twist angle is small.

On the other hand, when the twist angle approaches the magic value, the miniband bandwidth decreases rapidly \cite{key,TB2}, lattice relaxation and Coulomb interaction become important and can strongly renormalize the single-particle band dispersion. In particular, we envision that these correlation effects can significantly enhance the single-particle intervalley scattering  and hence the corresponding Dirac mass term, without breaking any lattice symmetry. If this scenario is correct, we expect the correlation-enhanced Dirac mass in TBG near charge neutrality will be controlled by the twist angle, becoming very small away from the magic angle. This prediction can be tested by systematically studying Landau level degeneracy in TBG with different twist angles.

In terms of our tight-binding model, the mass generation term corresponds to $ t_{1}' $ term in (\ref{TB2}) which break U(1) symmetry in orbital space. As we discussed in Section. III and shown in Fig. \ref{fig}a and b, for $t_{1}' \neq 0$, the unstable Dirac fermions are gapped out at low-energy and only half of Dirac fermions remain robust. As a result, the  Landau level degeneracy becomes 4-fold, coming from $\pm K$ points of MBZ and electron spin.

\section{Outlook}

The two-orbital honeycomb Hubbard model proposed in this work provides a theoretical  framework for studying correlated electron phenomena in graphene superlattices. Many important questions remain to be addressed, among which we highlight a few. First, the nature of the correlated insulator ground state at $q=1$. It may exhibit long-range order, such as spin, orbital or valence bond solid order. Alternatively, because of the proximity to metal-insulator transition and/or the presence of orbital fluctuations \cite{Zaanen}, the correlated insulator may be a quantum spin liquid. In this regard, it is worth noting that in the special limit of our Hubbard model with SU(4) symmetry, the effective Hamiltonian for the correlated insulator at strong coupling is the SU(4) generalization of Heisenberg model (Kugel-Khomskii model) on the honeycomb lattice. Analytical and numerical studies of this model indicate the lack of any long-range order \cite{Zhang}, and suggests a possible quantum spin liquid with gapless neutral fermionic excitations \cite{spinliquid}. Thermal conductivity measurements can tell the existence or absence of gapless neutral excitations in the correlated insulator state of TBG.

Our model also provides a starting point for studying superconductivity in TBG. The strong electron repulsion disfavors on-site pairing, and opens the possibility of unconventional pairing symmetry. We leave the study of superconductivity in the two-orbital honeycomb Hubbard model to a future work.

\section*{Acknowledgment}

We thank Pablo Jarillo-Herrero, Yuan Cao and Mikito Koshino for invaluable discussions. This work is supported by DOE Office of Basic Energy Sciences, Division of Materials Sciences and Engineering under Award DE-SC0010526. LF is partly supported by the David and Lucile Packard Foundation.

{\it Note added}: Recently the prediction of this work based on symmetry analysis that  Wannier orbitals of the lowest mini-bands of twisted bilayer graphene have  $(p_x, p_y)$ symmetry and form a honeycomb lattice has been verified by explicit numerical calculations \cite{Vafek,Koshino}. These works also found tight-binding models of the same form as ours. In contrast, a different honeycomb model was proposed in Ref. \cite{Senthil}, where orbitals have two different on-site energies, instead of being degenerate as dictated by the $D_3$ lattice symmetry of twisted bilayer graphene.

\appendix
\setcounter{figure}{0}
\renewcommand{\thefigure}{S\arabic{figure}}

\begin{widetext}
\section*{Supplementary Material for ``Model for Metal-Insulator Transition in Graphene Superlattices and Beyond"}

\section{Possible Tight-Binding Models}
In this section we list all possible tight-binding models and show that only honeycomb lattice with degenerate $p_{x,y}$ orbitals can reproduce the band symmetry at all high symmetry points.

First, when the centers of Wannier orbitals are not at high symmetry points of the superlattice, the number of lowest bands should be multiples of 6 due to the point group $ D_3 $, while we only have four lowest bands from numerical calculations. Thus the Wannier centers should be at high symmetry points under $ D_3=\{C_{2y}^mC_{3z}^n|m,n\in\mathbb{Z}\} $. When Wannier centers are along $ C_{2y} $ rotation axis but not at $ C_{3z} $ rotation invariant points (the so-called saddle points, SP for short), they form a kagome lattice, which contains 3 superlattice sites in each supercell. The number of lowest bands in kagome lattice will be multiples of 3, which is also not consistent with four lowest bands from numerical calculations.

When Wannier centers are not along $ C_{2y} $ rotation axis but at $ C_{3z} $ rotation invariant points (namely AB and BA spots), they will form a honeycomb lattice, whose number of lowest bands is multiples of 2. Notice that AB and BA spots are invariant under $ C_{3z} $ followed by a superlattice translation. When Wannier centers are both along $ C_{2y} $ rotation axis and at $ C_{3z} $ rotation center (namely AA spots), the triangular lattice is formed and the number of lowest bands can be any positive integer. The kagome, honeycomb and triangular lattices are shown in Fig. \ref{S0}.

\begin{figure}
\begin{center}
\leavevmode\includegraphics[width=01.0\hsize]{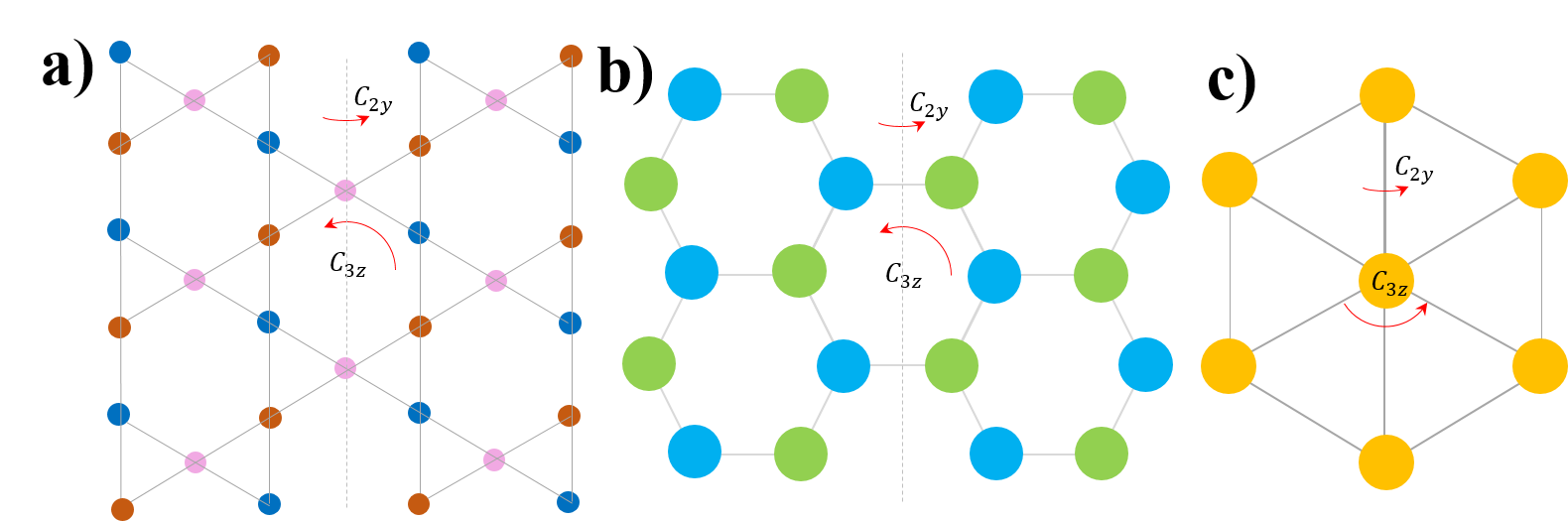}
\end{center}
\caption{Three lattices formed by high symmetry points of twisted bilayer graphene. They are a) kagome lattice by SP spots, b) honeycomb lattice by AB/BA spots and c) triangular lattice by AA spots.}
\label{S0}
\end{figure}

We summarize the analysis of honeycomb and triangular lattices in Table. \ref{S}. Since in $ D_3 $ point group the angular momentum is defined modulo 3, under $D_3$ higher order orbitals will transform the same as $ (p_x ,p_y) $ orbitals, $ s $ or $ p_z $ orbital, depending on the details of the orbitals. For example, $ (d_{x^2-y^2},d_{xy}) $ orbitals are the same as $ (p_x ,p_y) $ orbitals under $D_3$. In other words, the analysis of $ (p_x ,p_y) $ orbitals, $ s $ and $ p_z $ orbital covers all types of orbitals. For honeycomb lattice with $ (p_x ,p_y) $ orbitals the analysis is given in the maintext and the band symmetry at all high symmetry points can be correctly reproduced. Any other orbital or lattice will not reproduce the band symmetry at all high symmetry points as shown in Table. \ref{S}. Some combinations cannot reproduce $ \Gamma $ point while some fail at $K$ point.

For honeycomb lattice with $ s $ or $ p_z $ orbital, denote $ a_i $ as the annihilation operator at site $ i $, the Bloch state at momentum $\bm k$ can be constructed as
\begin{eqnarray}
a_{\bm k}=\sum_{i}e^{i\bm k\cdot\bm R_{i}}a_{i},
\end{eqnarray}
where $ \bm R_{i} $ is the coordinate of honeycomb lattice site $i$.
We find under $C_{3z}$, the Bloch state at $ \Gamma $ point transforms as
\begin{eqnarray}
a_{\bm\Gamma}\equiv\sum_{i}a_{i}\to
\sum_{i'}a_{i'}
=a_{\bm\Gamma},\quad
\bm R_{i'}=C_{3z}\bm R_{i}.
\end{eqnarray}
Hence the eigenstate at $ \Gamma $ point will be a singlet, which is not consistent with the doublet representation found in numerical band structure.

For triangular lattice with $ (p_x ,p_y) $ orbitals, the Bloch state is
\begin{eqnarray}
c_{\bm k,\tau}=\sum_{i}e^{i\bm k\cdot\bm R_{i}}c_{i\tau}
\end{eqnarray}
where $ \tau=\pm $ denotes the angular momentum and hence the type of Wannier orbital, $ \bm R_{i} $ is the coordinate of triangular lattice site $i$, $ c_{i\tau} $ annihilates an electron at site $i$ with orbital $\tau$. We find under $C_{3z}$, the Bloch state at $ +\bm K $ point transforms as
\begin{eqnarray}
c_{\bm K,\tau}\to 
\sum_{i}e^{i\bm K\cdot C_{3z}\bm R_{i}}e^{-2i\tau\pi/3}c_{i\tau}=\sum_{i}e^{i\bm K\cdot\bm R_{i}}e^{-2i\tau\pi/3}c_{i\tau}=e^{-2i\tau\pi/3}c_{\bm K,\tau}
\end{eqnarray}
Hence the eigenstates at $ \bm K $ point will be two doublets $ \{E,E\} $, which is not consistent with the representation $ \{A_1,A_2,E\} $ from microscopic considerations. Notice that at $ +\bm K $ point, the Bloch wave function in triangular lattice is invariant under $ C_{3z} $, while the Bloch wave function in honeycomb lattice has finite angular momentum $ L_z=\pm 1 $ under $ C_{3z} $, as shown in Fig. \ref{S1}.

\begin{figure}
\begin{center}
\leavevmode\includegraphics[width=01.0\hsize]{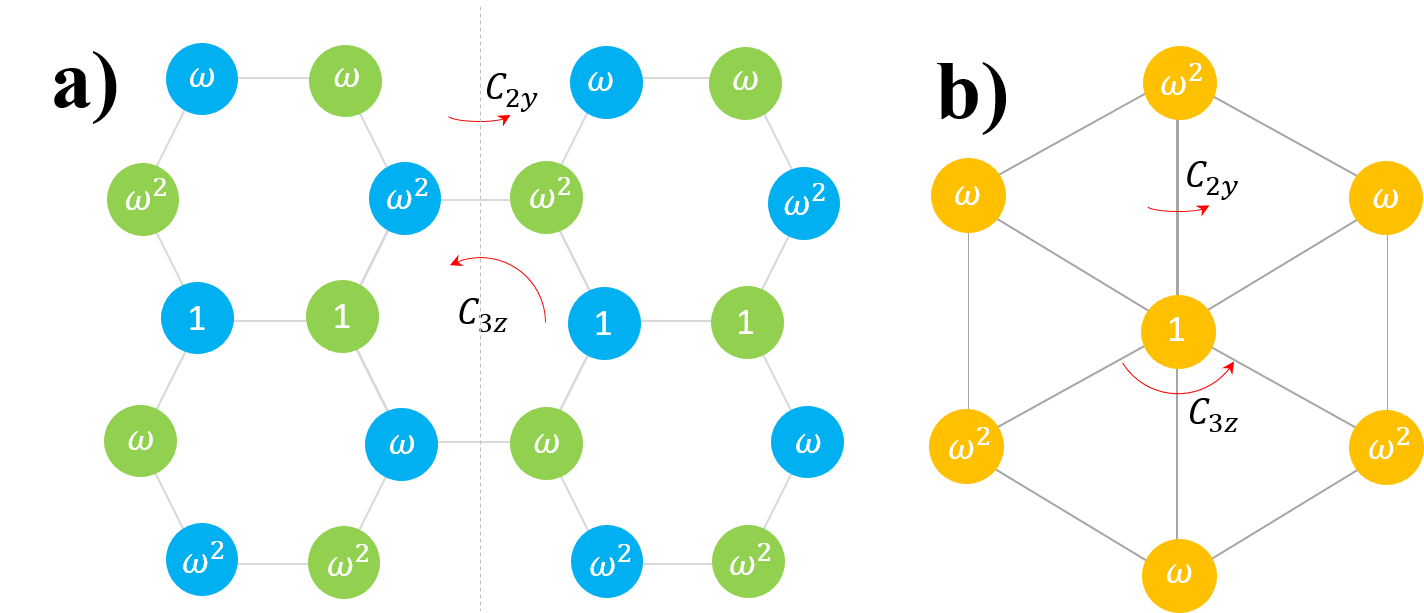}
\end{center}
\caption{Bloch wave function at $ +\bm K $ point for a) honeycomb and b) triangular lattices. Here only the phases due to Bloch wave are shown at lattice sites and the orbital parts are neglected, $ \omega =\exp(2\pi i/3) $. Rotations $ C_{3z},C_{2y} $ are also shown. It can be found that under $ C_{3z} $, the Bloch wave function of honeycomb lattice will have finite angular momentum $ L_{z}=\mp 1 $ for green and blue sublattices respectively, while the Bloch wave function of triangular lattice will have zero angular momentum $ L_{z}=0 $.}
\label{S1}
\end{figure}

For triangular lattice with $ s $ or $ p_z $ orbital, denote $ a_i $ as the annihilation operator at site $ i $, the Bloch state at momentum $\bm k$ can be constructed as
\begin{eqnarray}
a_{\bm k}=\sum_{i}e^{i\bm k\cdot\bm R_{i}}a_{i},
\end{eqnarray}
where $ \bm R_{i} $ is the coordinate of triangular lattice site $i$.
We find under $C_{3z}$, the Bloch state at $ \Gamma $ point transforms as
\begin{eqnarray}
a_{\bm\Gamma}\equiv\sum_{i}a_{i}\to
\sum_{i'}a_{i'}
=a_{\bm\Gamma},\quad
\bm R_{i'}=C_{3z}\bm R_{i}.
\end{eqnarray}
Hence the eigenstate at $ \Gamma $ point will be a singlet, which is not consistent with the doublet representation found in numerical band structure.

\begin{table}
\centering
\begin{tabular}{c|c|ccc} \hline
Lattice & Orbital   & $ \Gamma (D_3) $          & $ K (D_3) $          & $M(C_2)$     \\
            \hline
Honeycomb  & $ (p_x ,p_y),(d_{x^2-y^2},d_{xy}) $ & $ \{E,E\} $ & $\{A_{1},A_{2},E\}$   & $\{A,A,B,B\}$ \\
& $ s,p_{z},f_{y(3x^2-y^2)},f_{x(x^2-3y^2)} $   & $ \{A_1,A_2\} $ & $E$   & $\{A,B\}$ \\
\hline     
Triangular & $ (p_x ,p_y),(d_{x^2-y^2},d_{xy}) $ & $ E $ & $ E $   & $\{A,B\}$ \\
& $ s,f_{y(3x^2-y^2)} $   & $ A_1 $ & $ A_1 $   & $ A $ \\
& $ p_{z},f_{x(x^2-3y^2)} $   & $ A_2 $ & $ A_2 $   & $ B $ \\
\hline
\end{tabular}
\caption{Symmetries of eigenstates at $ \Gamma,K,M $ points in different tight-binding models.}
\label{S}
\end{table}

\section{Illustrative Examples of Eigenstates at $\Gamma$ Point}
Here we show illustrative examples of energy eigenstates $ \Psi_{e,h}^{\Gamma} $ at $\Gamma$ point with $L_{z}=-1,+1,0$. In all these examples there is accidental particle-hole symmetry and states at $ A $ sublattice always dominate.

We find that the lowest two eigenstates $ \Psi_{e,h}^{\Gamma} $ both have angular momentum $ L_z=+1 $ if we choose
\begin{eqnarray}
T_{0}=\frac{v_F}{5\sqrt{3}a}
\begin{pmatrix}
1&1\\
1&-0.6
\end{pmatrix},
T_{\pm}=\frac{v_F}{5\sqrt{3}a}
\begin{pmatrix}
1&\omega^{\pm}\\
\omega^{\pm}&0.6\omega^{\mp}
\end{pmatrix},
\end{eqnarray}
where $ \omega =e^{2i\pi/3} $. The eigenenergies and real-space probability distribution of $ \Psi_{e,h}^{\Gamma} $ are shown in Fig. \ref{2}. It can be found that
\begin{eqnarray}\label{ep}
\Psi^{\Gamma}_{e} &=& \alpha U^{A}(\bm R_{c})+\beta U^{B}(-\bm R_c)+\alpha^{*}L^{A}(-\bm R_{c})+\beta^{*}L^{B}(\bm R_c),\\
\Psi^{\Gamma}_{h} &=& \alpha U^{A}(\bm R_{c})+\beta U^{B}(-\bm R_c)-\alpha^{*}L^{A}(-\bm R_{c})-\beta^{*}L^{B}(\bm R_c)
\end{eqnarray}
where $ (|\alpha|,|\beta|)=(0.2770,0.1643),$ and $(\text{arg}\alpha,\text{arg}\beta)=(0.5250,0.0214) $. The eigenenergies and real-space probability distribution of $ \Psi_{e,h}^{\Gamma} $ of this case are shown in Fig. 4 of the maintext.

When we choose
\begin{eqnarray}
T_{0}=\frac{v_F}{5\sqrt{3}a}
\begin{pmatrix}
1&1\\
1&0.6
\end{pmatrix},
T_{\pm}=\frac{v_F}{5\sqrt{3}a}
\begin{pmatrix}
1&1\\
1&0.1
\end{pmatrix},
\end{eqnarray}
we find that the lowest two eigenstates $ \Psi_{e,h}^{\Gamma} $ both have angular momentum $ L_z=-1 $ and
\begin{eqnarray}\label{ee}
\Psi^{\Gamma}_{e} &=& \alpha U^{A}(-\bm R_{c})+\beta U^{B}(\bm 0)+\alpha^{*}L^{A}(\bm R_{c})+\beta^{*}L^{B}(\bm 0),\\
\Psi^{\Gamma}_{h} &=& \alpha U^{A}(-\bm R_{c})+\beta U^{B}(\bm 0)-\alpha^{*}L^{A}(\bm R_{c})-\beta^{*}L^{B}(\bm 0)
\end{eqnarray}
where $ (|\alpha|,|\beta|)=(0.2943,0.2065),$ and $(\text{arg}\alpha,\text{arg}\beta)=(0.1566,0.2446) $. 

We note that according to Eq.(\ref{e}), the component of Bloch wavefunction $\Psi^\Gamma$ on the $A$ sublattice has its maxima at AB spots on layer 1 and at BA spots on layer 2, while the component on the $B$ sublattice has its maxima around AA spots on both layers. Following the same symmetry argument in the maintext we can understand that maxima of $A$ sublattice component should be at AB and BA spots.
In contrast, $B$ sublattice sites are not at rotation centers, and the intra-unit-cell phase factors $e^{i {\bm K}_1 \cdot {\bm x}^B}$ and $e^{i {\bm K}_2 \cdot {\bm y}^B}$ already carry the angular momentum $ -1 $. As a result, the envelope function of B sublattice wavefunction carries zero angular momentum and is therefore allowed to be peaked at AA spots.

The maxima of envelope part of $\Psi_{e}^{\Gamma}$ and $\Psi^{\Gamma}_{h}$ form a honeycomb superlattice consisting of two layers of corrugated triangular superlattices in both cases of $ L_{z}=\pm 1 $. Around each maximal spot, the state is dominated by states from central $A$ sites and states from surrounding $B$ sites, resulting in a three-fold structure. When $ L_{z}=+1 $, the $AB$ bonds will be pointing away from three-fold rotation centers. When $ L_{z}=-1 $, the $AB$ bonds connecting probability centers between $ A $-site and $ B $-site states will be pointing towards three-fold rotation centers.

\begin{figure}
\begin{center}
\leavevmode\includegraphics[width=01.0\hsize]{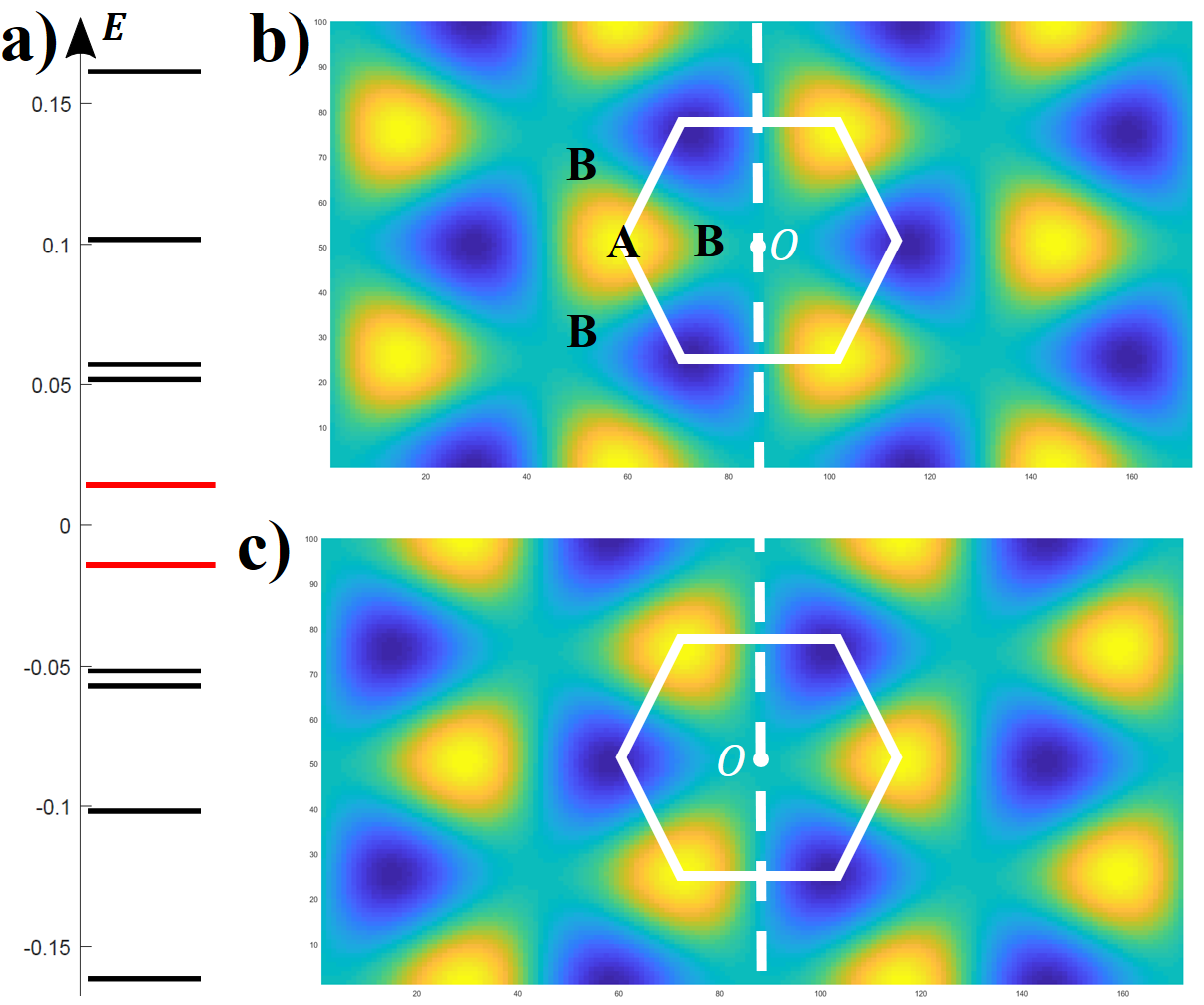}
\end{center}
\caption{Real space probability distributions $ |\Psi_{e}^{\Gamma}(\bm x)|^2=|\Psi_{h}^{\Gamma}(\bm x)|^2 $ of lowest eigenstates at $\Gamma$ point with angular momentum $ L_{z}=-1 $. Eigenenergies are shown in a) in unit of $2v_{F}/(\sqrt{3}a)$, where the lowest energies are in red. $ |\Psi_{e,h}^{\Gamma}(\bm x)|^2 $ in layer 1 and 2 are shown in b) and c) respectively. The hexagon is the unit cell of graphene superlattice and $ O $ is origin.}
\label{2}
\end{figure}

The case with $ L_{z}=0 $ can be realized with
\begin{eqnarray}
T_{0}=\frac{v_F}{5\sqrt{3}a}
\begin{pmatrix}
1&1\\
1&0.6
\end{pmatrix},\quad
T_{\pm}=\frac{v_F}{5\sqrt{3}a}
\begin{pmatrix}
0.1&1\\
1&1
\end{pmatrix},
\end{eqnarray}
and is shown in Fig. \ref{S2}. And the wavefunctions are found as following
\begin{eqnarray}\label{WF}
\Psi^{\Gamma}_e &=& \alpha U^{A}(\bm 0)+\beta U^{B}(\bm R_c)+\alpha^{*}L^{A}(\bm 0)+\beta^{*}L^{B}(-\bm R_c)\\
\Psi^{\Gamma}_h &=& \alpha U^{A}(\bm 0)+\beta U^{B}(\bm R_c)-\alpha^{*}L^{A}(\bm 0)-\beta^{*}L^{B}(-\bm R_c).
\end{eqnarray}
where $ (|\alpha|,|\beta|)=(0.2450,0.3949),$ and $(\text{arg}\alpha,\text{arg}\beta)=(0.2329,0.2094) $.

In the case of $L_{z}=0$, we find a triangular lattice is formed by envelope peaks at AA spots. Thus it is tempting to construct a two-orbital tight-binding model on the triangle lattice. However, it turns out that the band symmetries at high symmetry points cannot be correctly obtained in this case since we will obtain $s$ and $p_{z}$ orbitals at each site of the triangular lattice, and obstructions may be encountered in constructing tight-binding models.

\begin{figure}
\begin{center}
\leavevmode\includegraphics[width=01.0\hsize]{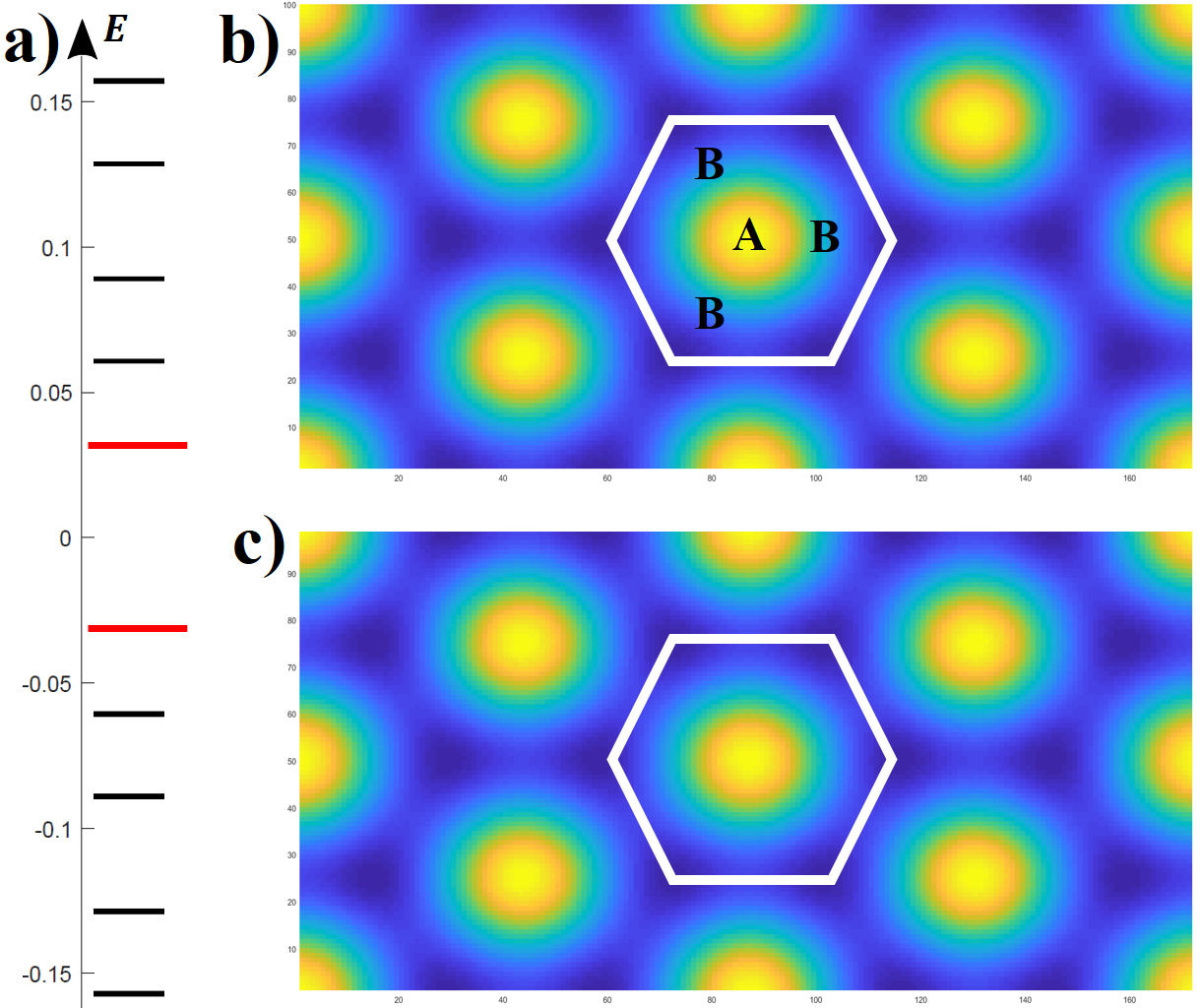}
\end{center}
\caption{Real space probability distribution $ |\Psi_{e}(\bm x)|^2=|\Psi_{h}(\bm x)|^2 $ of lowest eigenstates at $\Gamma$ point with angular momentum $ L_{z}=0 $ for both electron and hole sides. We consider states near $\bm K_{+}$ valley. Eigenenergies are shown in a) in unit of $2v_{F}/(\sqrt{3}a)$, where the lowest two are in red. Probability distributions in layer 1 and 2 are shown in b) and c) respectively.}
\label{S2}
\end{figure}

\end{widetext}


\begin{thebibliography}{99}
\bibitem{Cao1} Yuan Cao, Valla Fatemi, Ahmet Demir, Shiang Fang, Spencer L. Tomarken, Jason Y. Luo, J. D. Sanchez-Yamagishi, K. Watanabe, T. Taniguchi, E. Kaxiras, R. C. Ashoori and P. Jarillo-Herrero, Nature, doi:10.1038/nature26154 (2018).

\bibitem{Cao2} Yuan Cao, Valla Fatemi, Shiang Fang, Kenji Watanabe, Takashi Taniguchi, Efthimios Kaxiras and Pablo Jarillo-Herrero, Nature, doi:10.1038/nature26160 (2018).

\bibitem{Mele} E. J. Mele, Nature News and Views, doi:10.1038/d41586-018-02660-4.

\bibitem{Xu} Cenke Xu and Leon Balents, arXiv:1803.08057.

\bibitem{Volovik} G.E. Volovik, arXiv: 1803.08799 .

\bibitem{MacDonald} Rafi Bistritzer and Allan H. MacDonald, PNAS \textbf{108}, 12233 (2011).

\bibitem{MIT} Masatoshi Imada, Atsushi Fujimori, and Yoshinori Tokura, Rev. Mod. Phys. \textbf{70}, 1039 (1998).

\bibitem{Simons} J.P.F. LeBlanc, Andrey E. Antipov, Federico Becca, Ireneusz W. Bulik, Garnet Kin-Lic Chan, Chia-Min Chung, Youjin Deng, Michel Ferrero, Thomas M. Henderson, Carlos A. Jiménez-Hoyos, E. Kozik, Xuan-Wen Liu, Andrew J. Millis, N.V. Prokof’ev, Mingpu Qin, Gustavo E. Scuseria, Hao Shi, B. V. Svistunov, Luca F. Tocchio, I. S. Tupitsyn, Steven R. White, Shiwei Zhang, Bo-Xiao Zheng, Zhenyue Zhu, and Emanuel Gull (Simons Collaboration on the Many-Electron Problem), Phys. Rev. X \textbf{5}, 041041 (2015).

\bibitem{Stanford} S. Raghu, S. A. Kivelson, and D. J. Scalapino, Phys. Rev. B \textbf{81}, 224505 (2010).

\bibitem{Stripe} Bo-Xiao Zheng, Chia-Min Chung, Philippe Corboz, Georg Ehlers, Ming-Pu Qin, Reinhard M. Noack, Hao Shi, Steven R. White, Shiwei Zhang, Garnet Kin-Lic Chan, Science \textbf{358}, 1155 (2017).

\bibitem{key} Nguyen N. T. Nam and Mikito Koshino, Phys. Rev. B \textbf{96}, 075311 (2017).

\bibitem{TB1} Pilkyung Moon and Mikito Koshino, Phys. Rev. B \textbf{85}, 195458 (2012).

\bibitem{TB2} E. Suárez Morell, J. D. Correa, P. Vargas, M. Pacheco, and Z. Barticevic, Phys. Rev. B \textbf{82}, 121407(R) (2010).

\bibitem{TB3} G. Trambly de Laissardière, D. Mayou and L. Magaud, Nano Lett. \textbf{10}, 804 (2010).

\bibitem{TB4} G. Trambly de Laissardière, D. Mayou, and L. Magaud, Phys. Rev. B \textbf{86}, 125413 (2012).

\bibitem{TB5} A. O. Sboychakov, A. L. Rakhmanov, A. V. Rozhkov, and Franco Nori, Phys. Rev. B \textbf{92}, 075402 (2015).

\bibitem{DFT1} Shiang Fang and Efthimios Kaxiras, Phys. Rev. B \textbf{93}, 235153 (2016).

\bibitem{DFT2} Kazuyuki Uchida, Shinnosuke Furuya, Jun-Ichi Iwata, and Atsushi Oshiyama, Phys. Rev. B \textbf{90}, 155451 (2014).

\bibitem{DFT3} Sylvain Latil, Vincent Meunier, and Luc Henrard, Phys. Rev. B \textbf{76}, 201402(R) (2007).

\bibitem{Mele2} E. J. Mele, Phys. Rev. B \textbf{81}, 161405(R) (2010).

\bibitem{Neto} J. M. B. Lopes dos Santos, N. M. R. Peres, and A. H. Castro Neto, Phys. Rev. Lett. \textbf{99}, 256802 (2007).

\bibitem{Neto2} J. M. B. Lopes dos Santos, N. M. R. Peres, and A. H. Castro Neto, Phys. Rev. B \textbf{86}, 155449 (2012).

\bibitem{Koshino0} Pilkyung Moon and Mikito Koshino, Phys. Rev. B \textbf{87}, 205404 (2013).

\bibitem{SM} See Supplementary Material for eigenstates at $\Gamma$ point and consistency check for other types of tight-binding models.

\bibitem{LF} Liang Fu, Phys. Rev. Lett. \textbf{103}, 266801 (2009).

\bibitem{Cao3} Y. Cao, J. Y. Luo, V. Fatemi, S. Fang, J. D. Sanchez-Yamagishi, K. Watanabe, T. Taniguchi, E. Kaxiras, and P. Jarillo-Herrero, Phys. Rev. Lett. \textbf{117}, 116804 (2016).

\bibitem{George} Serge Florens and Antoine Georges, Phys. Rev. B \textbf{70}, 035114 (2004).

\bibitem{Kotliar} S. Florens, A. Georges, G. Kotliar, and O. Parcollet, Phys. Rev. B \textbf{66}, 205102 (2002).

\bibitem{badmetal} Edward Perepelitsky, Andrew Galatas, Jernej Mravlje, Rok Žitko, Ehsan Khatami, B. Sriram Shastry, and Antoine Georges, Phys. Rev. B \textbf{94}, 235115 (2016).

\bibitem{Zaanen} Louis Felix Feiner, Andrzej M. Oleś, and Jan Zaanen, Phys. Rev. Lett. \textbf{78}, 2799 (1997).

\bibitem{Zhang} Y.Q. Li, M. Ma, D.N. Shi, and F.C. Zhang, Phys. Rev. Lett. \textbf{81}, 3527 (1998).

\bibitem{spinliquid} P. Corboz, M. Lajkó, A. M. Lauchli, K. Penc, F. Mila, Phys. Rev. X \textbf{2}, 041013 (2012).


\bibitem{Vafek}
Jian Kang and Oskar Vafek, arXiv:1805.04918.

\bibitem{Koshino}
Mikito Koshino, Noah F. Q. Yuan, Takashi Koretsune, Masayuki Ochi, Kazuhiko Kuroki, Liang Fu, arXiv:1805.06819.

\bibitem{Senthil}
Hoi Chun Po, Liujun Zou, Ashvin Vishwanath, T. Senthil, arXiv:1803.09742.


\end{thebibliography}
\end{document}